\documentclass[letterpaper,11pt]{article}
\usepackage{deauthor}
\usepackage{times}
\usepackage{graphicx} 
\usepackage{amssymb}	%
\usepackage{amsmath}

\usepackage{xcolor,colortbl}    %

\usepackage{xspace}            %

\newcommand{\queryvis}{\textsf{QueryVis}\xspace}
\newcommand{\diagrams}{\textsf{Relational Diagrams}\xspace}

\newcommand{\sql}[1]{\textup{\textsf{\small#1}}}

\RequirePackage{scalerel}
\RequirePackage{tikz}

\usetikzlibrary{svg.path}

\definecolor{orcidlogocol}{HTML}{A6CE39}
\tikzset{
  orcidlogo/.pic={
    \fill[orcidlogocol] svg{M256,128c0,70.7-57.3,128-128,128C57.3,256,0,198.7,0,128C0,57.3,57.3,0,128,0C198.7,0,256,57.3,256,128z};
    \fill[white] svg{M86.3,186.2H70.9V79.1h15.4v48.4V186.2z}
                 svg{M108.9,79.1h41.6c39.6,0,57,28.3,57,53.6c0,27.5-21.5,53.6-56.8,53.6h-41.8V79.1z M124.3,172.4h24.5c34.9,0,42.9-26.5,42.9-39.7c0-21.5-13.7-39.7-43.7-39.7h-23.7V172.4z}
                 svg{M88.7,56.8c0,5.5-4.5,10.1-10.1,10.1c-5.6,0-10.1-4.6-10.1-10.1c0-5.6,4.5-10.1,10.1-10.1C84.2,46.7,88.7,51.3,88.7,56.8z};
  }
}

\DeclareRobustCommand{\orcidiconlink}[2]{%
    \hypersetup{urlcolor=black}%
    \href{https://orcid.org/#2}{#1 \mbox{\scalerel*{%
        \begin{tikzpicture}[yscale=-1,transform shape]%
            \pic{orcidlogo};%
        \end{tikzpicture}%
    }{|}}}%
}%

\usepackage[caption=false]{subfig} 

\newcommand{\introparagraph}[1]{\textbf{#1.}}        %

\newtoks\bsubfloattoks
\newdimen\bsubfloatht

\makeatletter
\newenvironment{bsubfloatrows}[1][\quad]
  {\def\bsubfloatspace{#1}\resetbsubfloatrows
   \def\\{\printbsubfloatrow\resetbsubfloatrows\par
     \@ifnextchar[{\bsubfloatvspace}{}}%
   \def\bsubfloatvspace[##1]{\vspace{##1}}%
  }
  {\printbsubfloatrow}
\newcommand{\bsubfloat}[2][]{%
  \sbox\z@{#2}%
  \ifdim\bsubfloatht<\ht\z@
    \bsubfloatht=\ht\z@
  \fi
  \bsubfloattoks=\expandafter{\the\bsubfloattoks
    \bsubfloatspace\subfloat[#1]{\vbox to\bsubfloatht{\hbox{#2}\vfill}}}%
}
\newcommand\resetbsubfloatrows{\bsubfloatht\z@\bsubfloattoks={\@gobble}}
\newcommand{\printbsubfloatrow}{\the\bsubfloattoks}
\makeatother

\usepackage{url}
\newtheorem{scenario}{Scenario}              	
\newtheorem{definition}{Definition}    %

\usepackage[colorlinks, allcolors=blue]{hyperref}

\newcommand{\osfpreregplain}{osf.io/mycr2}

\newcommand{\osfprereg}{\url{https://\osfpreregplain}}

\begin{document}

\title{Principles of Query Visualization}

\author{
  \hspace{-15mm}\phantom{x}
  \texorpdfstring{\orcidiconlink{Wolfgang Gatterbauer}{0000-0002-9614-0504}}{Wolfgang Gatterbauer}
  \hspace{-15mm}\phantom{x}  
  \\
  \hspace{-6mm}\phantom{x}
  Northeastern University
  \hspace{-6mm}\phantom{x}
  \\
  \footnotesize
  \hspace{-15mm}\phantom{x}
  w.gatterbauer@northeastern.edu
  \hspace{-15mm}\phantom{x}  
\and
  \texorpdfstring{\orcidiconlink{Cody Dunne}{0000-0002-1609-9776}}{Cody Dunne}\\
  \hspace{-6mm}\phantom{x}
  Northeastern University
  \hspace{-6mm}\phantom{x}
  \\
  \footnotesize
  \hspace{-15mm}\phantom{x}
  c.dunne@northeastern.edu
  \hspace{-15mm}\phantom{x}  
\and
  \texorpdfstring{\orcidiconlink{H.V.\ Jagadish}{0000-0003-0724-5214}}{H.V.\ Jagadish}\\
  \hspace{-6mm}\phantom{x}
  University of Michigan
  \hspace{-6mm}\phantom{x}
  \\
  \footnotesize
  jag@umich.edu
\and
  \hspace{-15mm}\phantom{x}
  \texorpdfstring{\orcidiconlink{Mirek Riedewald}{0000-0002-6102-7472}}{Mirek Riedewald}
  \hspace{-15mm}\phantom{x}  
  \\
  \hspace{-6mm}\phantom{x}  
  Northeastern University
  \hspace{-6mm}\phantom{x}
  \\
  \footnotesize
  \hspace{-15mm}\phantom{x}  
  m.riedewald@northeastern.edu  
  \hspace{-15mm}\phantom{x}  
}

\maketitle

\begin{abstract}
Query Visualization (QV) is the problem of transforming a given query into a graphical representation
that helps humans understand its meaning.
This task is notably different from designing a Visual Query Language (VQL)
that helps a user compose a query.
This article discusses 
the principles of relational query visualization
and its potential for simplifying user interactions with relational data.

\end{abstract}

\section{What is Query Visualization (QV) and what is it for?} 
\label{sec:1}

\begin{figure}
    \centering
    \begin{minipage}{0.5\textwidth}
        \centering
        \includegraphics[scale=0.425]{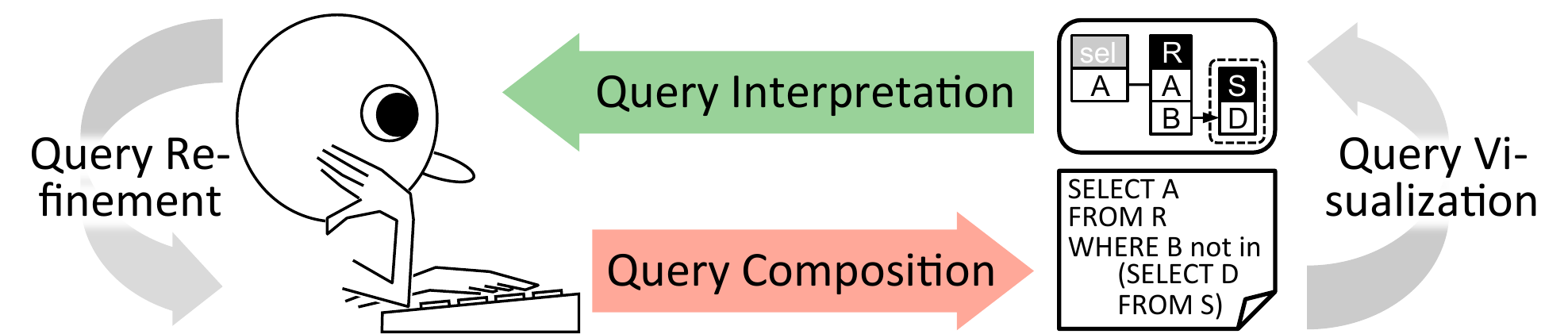}
		\caption{The goal of query visualization is to complement (but not substitute) the composition of queries 
		by creating automatic visualizations of queries. 
		Composition of a query is still performed via unambiguous and expressive text. 
		The transformation from text into a visualization can abstract away from a concrete syntax and thus be non-injective. 
		Compare to the write-format-preview cycle used by LaTeX~\cite{latex} in that a user writes text, 
		the system then autoformats and renders a document, 
		which the user can then peview.
		}
		\label{Fig_TheVision}
    \end{minipage}
	\hfill
    \begin{minipage}{0.45\textwidth}
        \centering
        \includegraphics[scale=0.36]{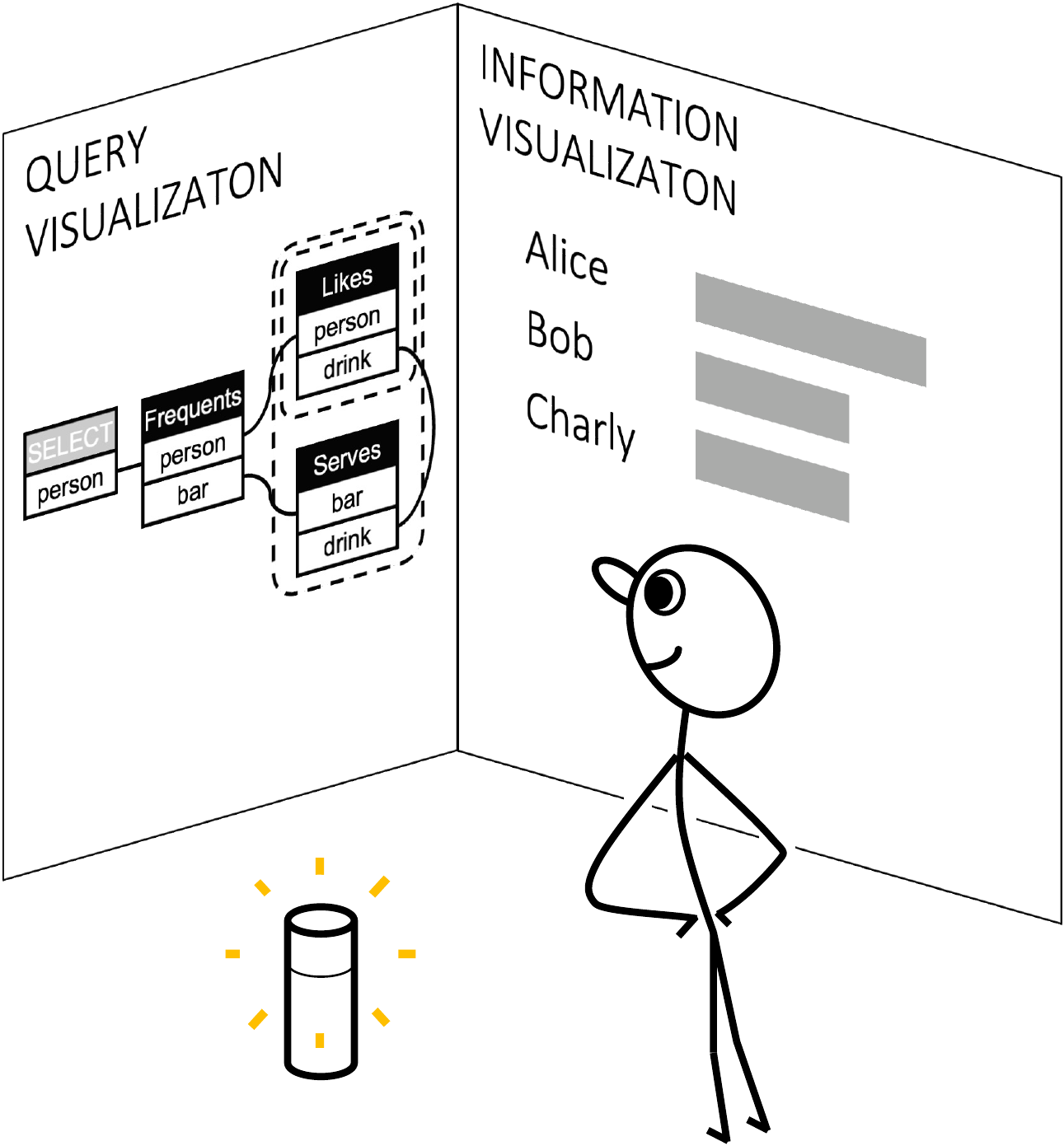}
		\caption{\autoref{scenario:5}: An analyst dictates queries to her voice assistant which then shows the query as understood 
		together with the query answers.}
		\label{Fig_SpeechAssistant}
    \end{minipage}
\end{figure}

The design of relational query languages and the difficulty for users to compose relational queries have
received much attention over the 
last 40 years~\cite{DBLP:journals/vlc/CatarciCLB97, 
ChanUserDatabaseInterface:1993,
GREENE1990303,
Harel:Nonprocedural:1985,
FrameworkForChoosingQueryLanguages:1985,
LEGGETT1984493,
DBLP:journals/csur/Reisner81, 
Reisner1975:HumanFactors,
Welty-Stemple:1981,
scamell:1993}.
A complementary and much-less-studied problem is that of helping users 
\emph{read and understand an existing query}.  
Reading code is hard, and SQL is no exception.  
With the proliferation of public data sources, and associated queries, users increasingly have a need to read other people's queries and scripts.  
Furthermore, it is usually much easier to modify a draft than to write something from scratch.  
As such, modifying an already existing query could be an effective way to write new queries.
However, modifying an existing query requires first to understand it (\autoref{Fig_TheVision}). 
For these reasons, it is valuable to help users understand queries,
and visualization is one obvious route.
In this paper, we study the problem of query visualization with a view towards improving query understanding.

\begin{figure}[t]
   \centering
\subfloat[SQL query]{
	\includegraphics[scale=0.58]{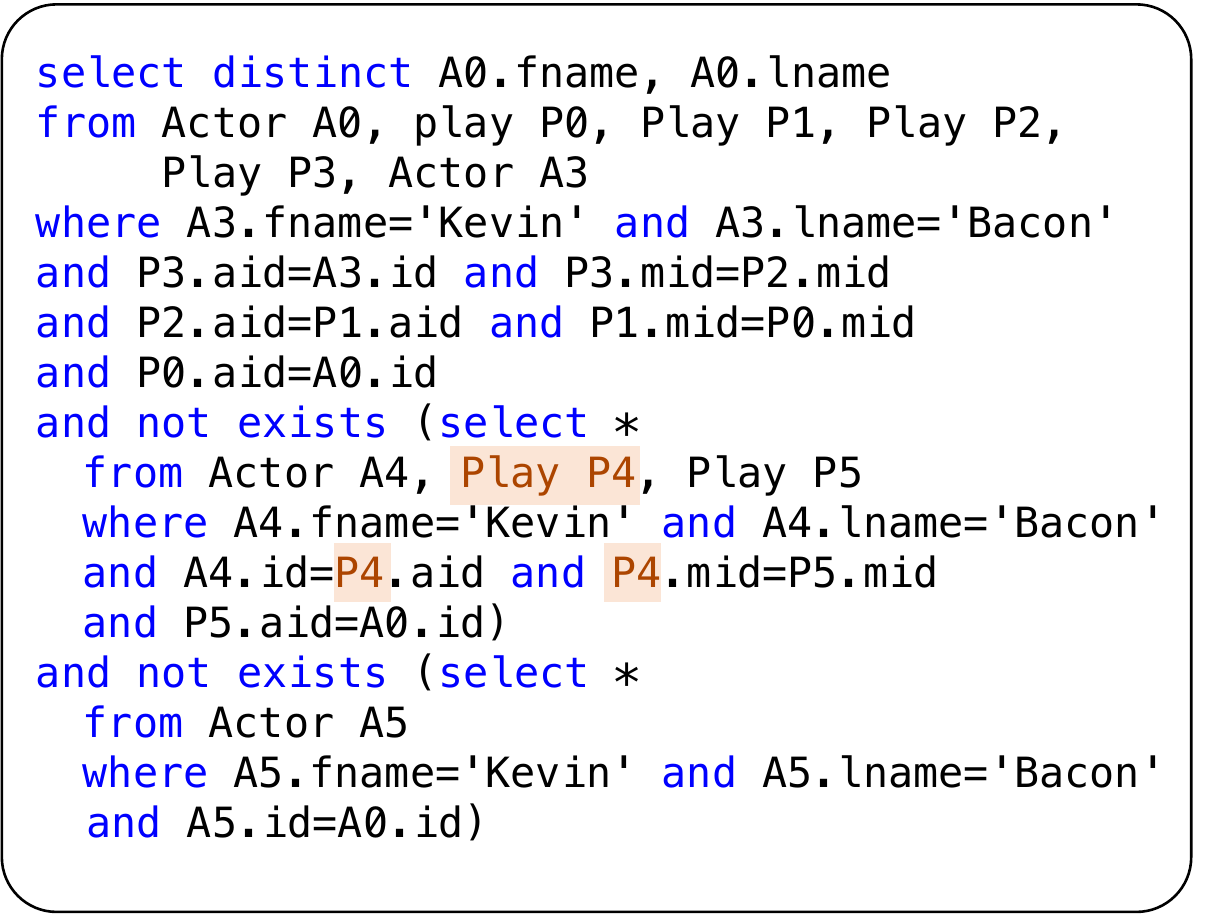}
	\label{Fig_KevinBacon_a}}
\hspace{4mm}
\subfloat[Query Visualization]{
	\includegraphics[scale=0.58]{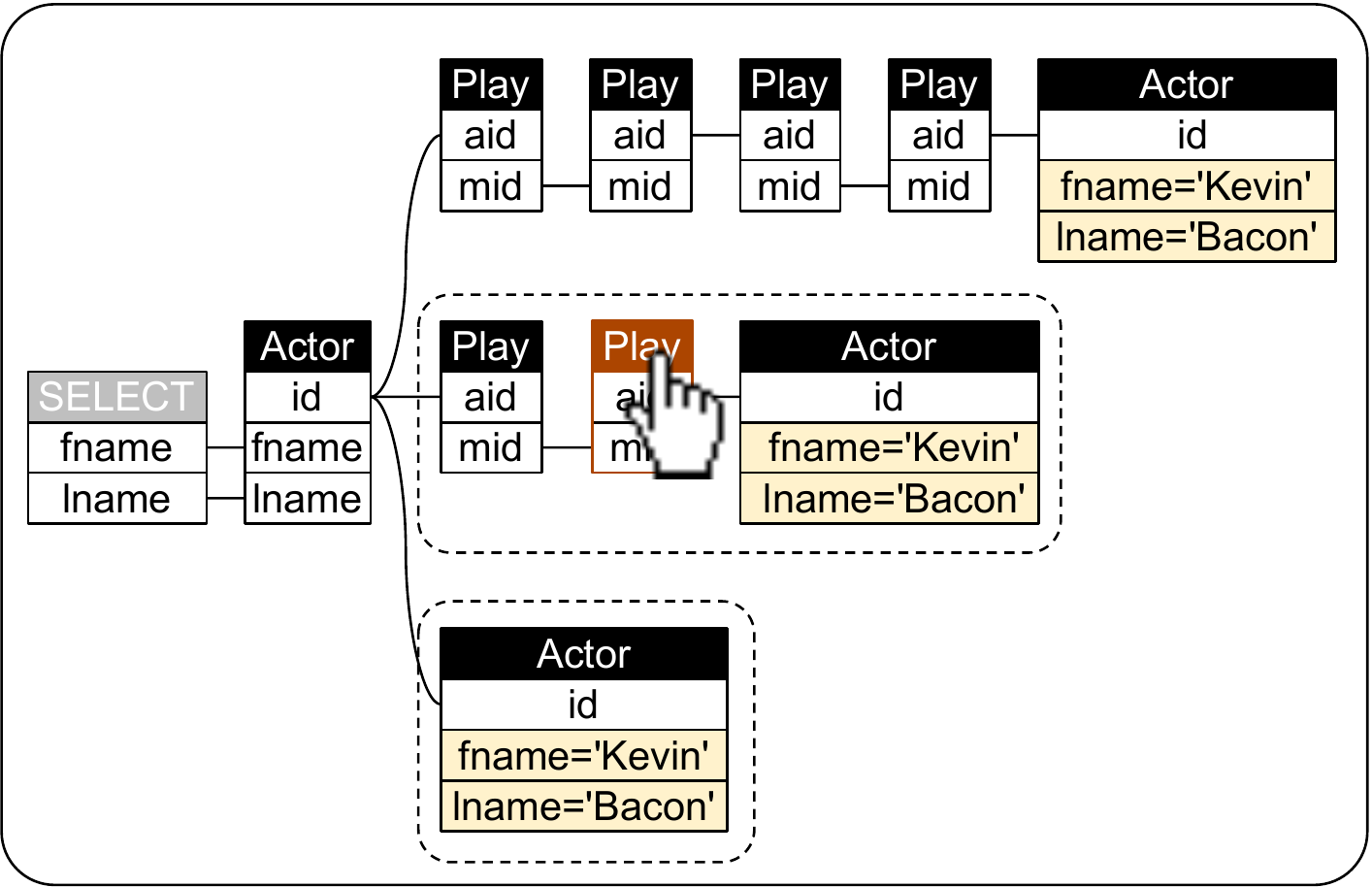}
	\label{Fig_KevinBacon_b}}	
	\caption{\autoref{scenario:1}: A user searches for a query by browsing through a repository 
	of previously-recorded SQL queries.
	For each query, she needs to \emph{quickly understand} its meaning. 
	A query visualization panel helps her understand the query by showing a 
	succinct representation of its relational query pattern. 
	The shown query returns actors with Bacon number 2.
	As she hovers her mouse over parts of the query, 
	both the textual and visualized query highlight corresponding parts in synchronization.
}
\label{Fig_KevinBacon}
\end{figure}

Consider the following five scenarios that illustrate how query visualizations can assist users:

\begin{scenario}[Data scientists reusing past queries]\label{scenario:1}
	A group of data analysts 
	are collaboratively analyzing movie data.
	This data is stored in a shared data repository. 
	In addition to the data itself, they are also sharing their queries using a Collaborative Query Management System (CQMS)
	\cite{QueRIERecommendations:2010,
	ARZAMASOVA2021101646,
	DBLP:conf/ssdbm/ChatzopoulouEP09,
	Eirinaki:QueRie:2014,
	DBLP:conf/icde/FanLZ11,
	HoweC2010:SQLshare, SQLshare:2016,
	DBLP:conf/cidr/KhoussainovaBGKS09, KhoussainovaKBS:2011, LiFWWF2011:DBease,
	Marcel:QueryRecommendations:2011,
	Milo:REACT:2016}.	
	The query recommendation component of that tool suggests relevant previously-issued queries that the user can choose from, 
	rather than write a query from scratch.
	The tool shows the queries both in text (\autoref{Fig_KevinBacon_a})
	and as query visualization (\autoref{Fig_KevinBacon_b}).
	The visualization preserves the logical structure of the textual query.
	 There is also a one-to-one mapping between the query and its visualization:
	As the user moves the mouse over components of the visualization, the corresponding component in the textual query is highlighted (and v.v.).
	The particular query used in 
	\autoref{Fig_KevinBacon}
	is over the IMDB movie database and finds all actors with Kevin Bacon number 2
	(i.e.\
	actors who have not played in a movie with Kevin Bacon directly, 
	but who have played with other actors who have played with Kevin Bacon).\footnote{See \url{https://youtu.be/kVFnQRGAQls?t=170} for an animated explanation of how to read and understand  this particular query.}
	These diagrams require some training to understand (see our discussion in \autoref{sec:2} for how to read them). 
	However, in a controlled and pre-registered user study  (see \autoref{sec:userstudy})
	we found that even a few minutes of training
	suffice, and experienced SQL users could interpret queries \emph{in less time} and \emph{with fewer errors} 
	using these diagrams instead of using SQL alone \cite{DBLP:conf/sigmod/LeventidisZDGJR20}.

	A widely known example of such a shared data and query repository is the Sloan Digital Sky Survey (SDSS)~\cite{QueRIERecommendations:2010, SDSS}.
	Data about stars was put into a relational database and is freely available for access. 
	Astronomers, who are mostly ``hobbyist" SQL users, 
	have to write queries to get the data they want.
	In most cases, the queries they wish to write are similar to queries others have written before.  So the standard workflow is for the user 
	to look at previously issued queries, find one that is close to what they want, and then modify it to suit their analysis needs.
	In response, SDSS has added templates for commonly written query types.\footnote{\url{http://skyserver.sdss.org/dr8/en/help/docs/realquery.asp}} 

\end{scenario}

\begin{scenario}[Visual feedback during query editing]
	As the user starts editing the SQL query (\autoref{Fig_KevinBacon_a}), the query visualization gets updated too
	(\autoref{Fig_KevinBacon_b}, updates are not shown).
	When a syntactically-correct query does not give the expected result,
	the query visualization can help the user understand that an incorrect join pattern was used between the various subqueries. 
\end{scenario}

\begin{scenario}[Learning from galleries of relational query patterns]
	A data scientist wants to issue another query and looks for inspiration in a \emph{web gallery of SQL design patterns}.
	Similar to the way users of Matplotlib\footnote{\url{https://matplotlib.org/stable/gallery/index.html}},
	 D3\footnote{\url{https://d3-graph-gallery.com/}}, 
	and Altair\footnote{\url{https://altair-viz.github.io/gallery/index.html}} 
	program new visualizations by browsing through, copying from, and adapting existing designs~\cite{10.1145/3503490}, 
	such galleries enhance the technical skills of data scientists and learners by 
	showing a range of possible relational patterns and design templates to learn from 
	that would be hard to browse and make sense of 
	based on text alone.
	Similar programmer behaviors are found outside of visualization, where existing code templates, examples, and idioms are extensively copied and adapted.
	From IDE (Integrated Development Environment) logs of 81 developers, 
	Ciborowska et al.\ \cite{Ciborowska2018} identified many cases of opportunistic code reuse from the Web
	followed by editing the code. 
	LaToza et al.~\cite{Latoza2006} surveyed 157 programmers, 
	and 56\% agreed that understanding code that someone else wrote is a serious problem.
	Yang et al.~\cite{Yang2017} found that many blocks of Python code are copied from Stack Overflow 
	into open-source projects with slight modifications. 
	Ahmed et al.~\cite{Ahmed2015} found that 24\%
	of copy-and-paste events 
	among 21,770 users of Eclipse were from sources external to the IDE, though this is likely an overestimate. 
	Brandt et al.~\cite{Brandt2009} found that such copy-and-paste programming is particularly beneficial for programmers working in new domains:
	In their study of students learning to use a new framework, one-third of the participants' code consisted of modified versions of examples from the documentation. 
\end{scenario}

\begin{scenario}[Clustering pattern-identical queries]
	A teacher receives the SQL solutions for a homework from her 50 students. 
	An automatic correction tool, such as
	ADUSA~\cite{DBLP:conf/kbse/KhalekELK08},
	Cosette~\cite{DBLP:conf/cidr/ChuWWC17},
	Qex~\cite{DBLP:conf/lpar/VeanesTH10}
	TATest~\cite{MRJ:ExplainingWrongQueries:2019},
	or XData~\cite{DBLP:journals/vldb/ChandraCKRS015},
	determines that 40 of those solutions are correct. But those correct solutions ``look'' very different,
	even after applying some standard SQL pretty printer,
	such as sqlparse~\cite{sqlparse}.
	The queries use different table aliases, nesting patterns, join sequences, and at times different syntactic constructs, 
	such as
	implicit joins in the WHERE clause
	or infixed SQL-92 join notation.
	The teacher would like to cluster the 40 correct solutions not by their syntactic variants, but 
	by whether there
	are `truly' 
	\emph{novel patterns} beyond the 3 she currently knows.
	Query visualization makes it easier to cluster and compare the 40 queries, because a good visualization
	captures the essence of the query structure, abstracting away superficial
	syntactic differences.	
	And, indeed, the teacher finds 2 new patterns.
	She can now show and discuss 5 total patterns with the learners in the next class.
\end{scenario}

\begin{scenario}[Visual feedback from voice assistants]
	\label{scenario:5}
	We now switch to the year 2045.
	A data analyst 
	stands in her office analyzing some company data. 
	She directs possible queries to her voice assistant 
	which then visualizes on the walls the queries 
	together with the data (\autoref{Fig_SpeechAssistant}). 
	The visualization of the query provides immediate feedback on what the assistant understood.
\end{scenario}

Common to these 5 scenarios is that query visualization helps the users achieve new functionalities 
or increased efficiency in composing queries
as a \emph{complement} to query composition and 
\emph{not a substitute} for it.

\begin{definition}[Query Visualization]
The term ``query visualization'' refers to both ($i$) a graphical representation of a query 
and ($ii$) the process of transforming a given query into a graphical representation.
The goal of query visualization is to help users more quickly understand the intent of a query,
as well as its relational query pattern.
\end{definition}

When we say ``query visualization", we will typically mean the end result. 
From context, it should be clear to the reader on the few occasions when we mean the process rather than the end result.

\section{What Query Visualization is not}
\label{section:whatisQVnot}

\subsection{Query Visualization is not the same as a Visual Query Language (VQL)}
\label{QVisnotVQL}

Visual Query Languages (VQLs) provide languages to express queries in a visual format.
Visual Query Systems (VQSs) implement VQLs and generate queries from visual representations constructed by users~\cite{DBLP:reference/db/Catarci18a}.
Such visual methods for specifying relational queries have been studied
extensively
(a 1997 survey by Catarci et al.~\cite{DBLP:journals/vlc/CatarciCLB97} 
cites over 150 references), and
many commercial database products offer some visual interface for users to write SQL queries.  
In parallel, there is a centuries-old history on the study of formal diagrammatic reasoning systems \cite{DBLP:conf/iccs/Howse08}
with the goal of helping humans to reason in terms of logical statements.\footnote{A relational query is a logical formula with free variables. 
A logical statement has no free variables and is intuitively the same as a Boolean query that returns a truth value of TRUE or FALSE. }
Yet despite their extensive study and intuitive appeal, successful visual tools today mostly only 
\emph{complement instead of 
replace}
text for specifying queries.
Why has visual query specification not yet replaced textual query specification?

\begin{figure}[t]
\centering
\begin{minipage}{0.50\textwidth}
	\includegraphics[scale=0.35]{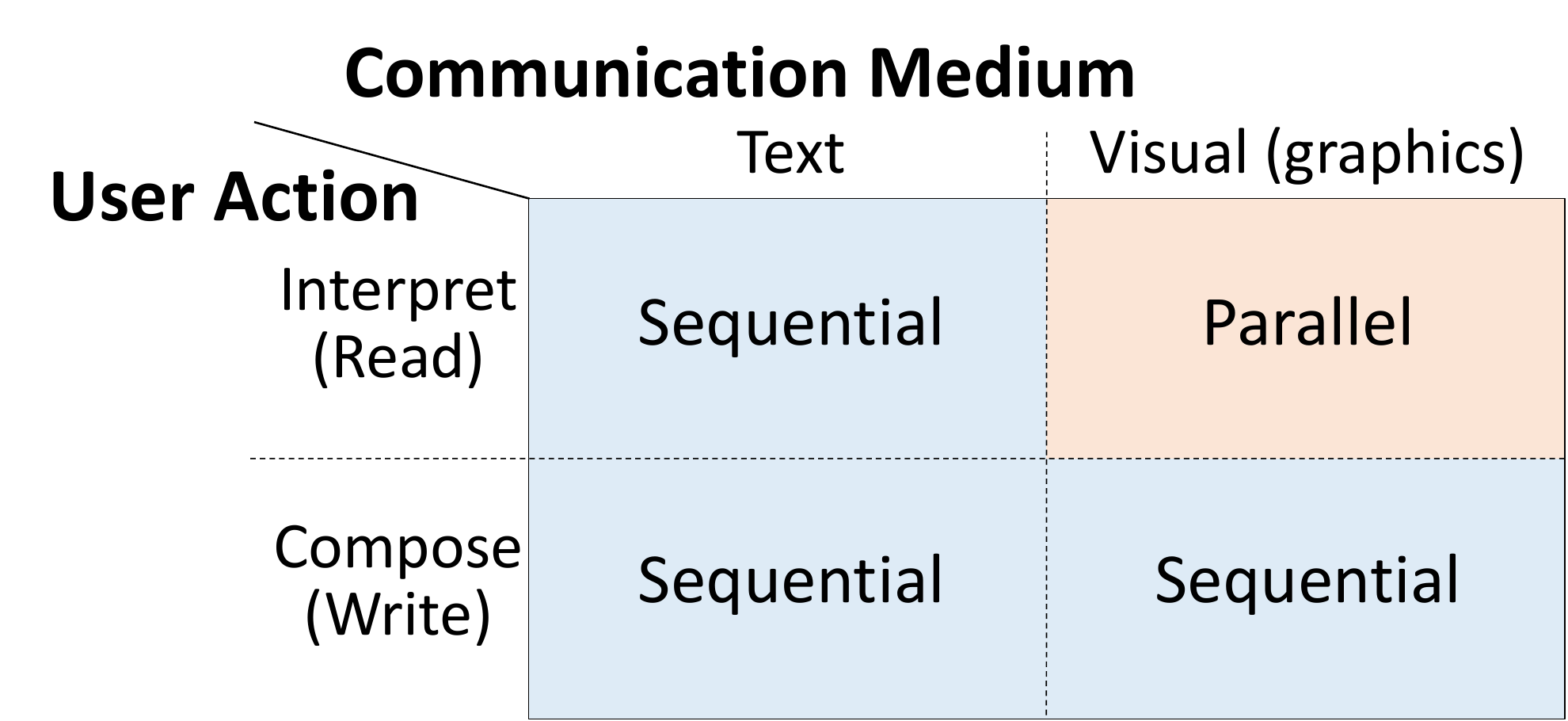}
	\caption{%
	    \emph{Composing} a query with a visual query language is as sequential as composing it with SQL. 
        \emph{Interpreting} a visualization (whether of information or a query) 
		is the only modus in which a user can act on information in parallel, 
		leveraging the speed of the human perceptual system 
		(orange = easier, blue = harder).
    }
    \label{Fig_MatrixDataQueryNew}
\end{minipage}
\hfill
\begin{minipage}{0.45\textwidth}
	\includegraphics[scale=0.35]{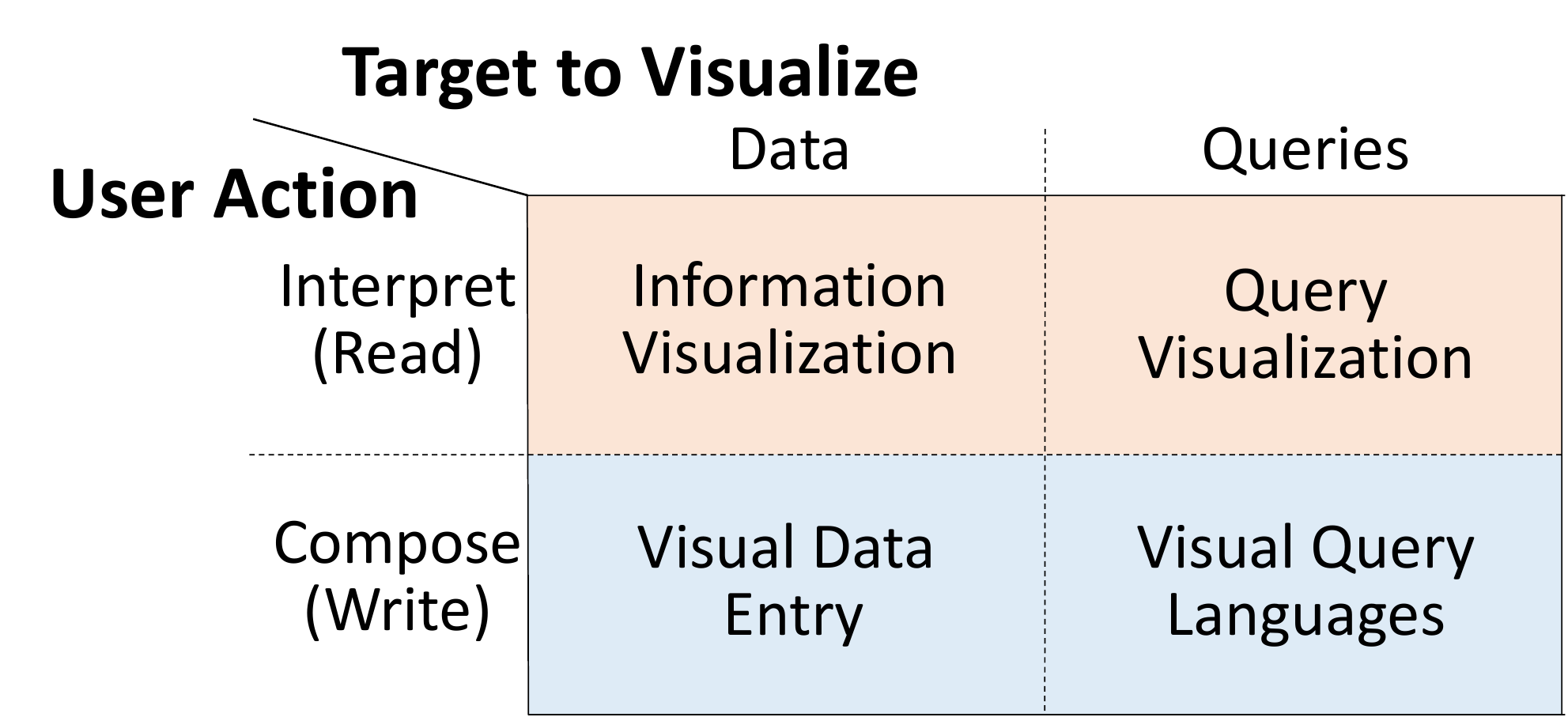}
	\caption{%
	    \emph{Visual Query Languages} allow a user to compose queries. 
		They have been widely studied and have a rich history.
	    In contrast, \emph{Query Visualization} helps the user understand an existing query 
		just as \emph{Information Visualization} helps understand data 
		(orange = easier, blue = harder).
    }
	\label{Fig_MatrixTextGraphics}
\end{minipage}
\end{figure}

We believe that there are two primary reasons:
(1) First, humans are \emph{better in interpreting rather than composing visuals} because visual composition is an inherently sequential process (\autoref{Fig_MatrixDataQueryNew}).
All human input methods (composition) are sequential, whether resulting in text or a graphic. Visual perception is a remarkable human sense (interpretation) that can 
understand inputs
in parallel,
and it works dominantly by \emph{spatial arrangement of information}. 
While reading text is also a visual activity, the spatial arrangement of the letters requires a sequential scan of the text 
(though notice that pretty printers can spatially arrange text, \autoref{section:alternatives}).
Hence, visual interpretation of graphics is the fastest way to communicate with humans, 
and it only works well for understanding rather than composing.
Even in theory, there is no dramatic speed-up 
in
using a visual language for composition. 
In practice, the user interaction is quite cumbersome: 
users must be able to interactively construct and manipulate expressions in a visual language and connect graphical elements 
to establish graphical relationships. In turn, the program must provide appropriate interpretations
 of mouse, touch, and keyboard events, 
 and it is difficult to build formal grammars and compilers for two-dimensional drawing areas. 
In sum, solutions to these graphical requirements are intricate, inherently difficult to implement, and challenging to use~\cite{Zhang:2007zr}. 

(2) A second reason is that graphs are more ambiguous than text, i.e.\ it is more difficult to be precise with a visual representation than with text. 
In order to precisely specify a query, possible options
and specific details affecting query semantics must be presented. 
In contrast, 
understanding a query requires a focus on the high-level structure, abstracting
away low-level details and subtleties. 
In programming languages, this distinction is clearly made between \emph{visual programming} for developing a program
and \emph{program visualization} for analyzing an existing program~\cite{MYERS199097}.

This leads us to suggest a user-query interaction 
that separates the query composition from the visualization (\autoref{Fig_TheVision}):
Composition is unchanged and best done in text 
(or alternatively with exploratory input
formats like natural language).
But composition is augmented and \emph{complemented with a visual that helps interpretation}. 
Recall \autoref{scenario:5} where a digital voice assistant connects to omnipresent screens to show what it understood before executing a command (\autoref{Fig_SpeechAssistant}).
Compare to the way that many of us write research papers: we write LaTeX code in a text editor but prefer to read and verify the auto-compiled PDF using automatic or editor-specific build instructions (\autoref{Fig_TheVision}).

Finally notice that query visualization is related to \emph{Information Visualization}~\cite{Chen:2006vn},
which also focuses on helping users understand complex relationships, 
but in data instead of in query logic~(\autoref{Fig_MatrixTextGraphics}).

\subsection{Query Visualization is not the same as Query Plan Visualization}
\label{sec:queryplans}

\begin{figure}[tb]
    \centering
	\subfloat[Postgres EXPLAIN command for $Q_{\textrm{some}}$]{
	\includegraphics[scale=0.5]{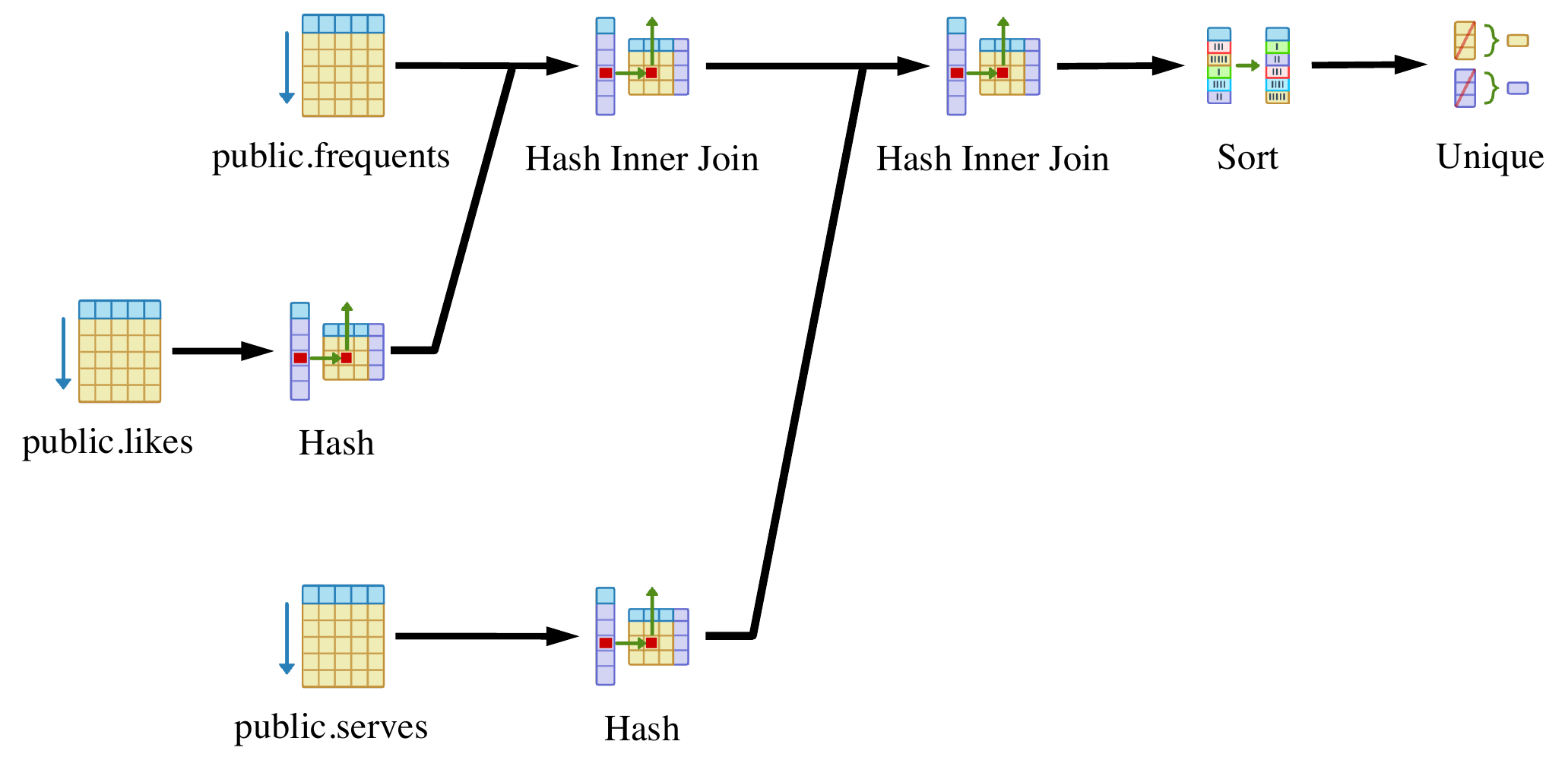}
	    \label{Fig_explain_PersonBarDrink}
	}
	\hfill
	\subfloat[$Q_{\textrm{some}}$ in DFQL]{
	\includegraphics[scale=0.5]{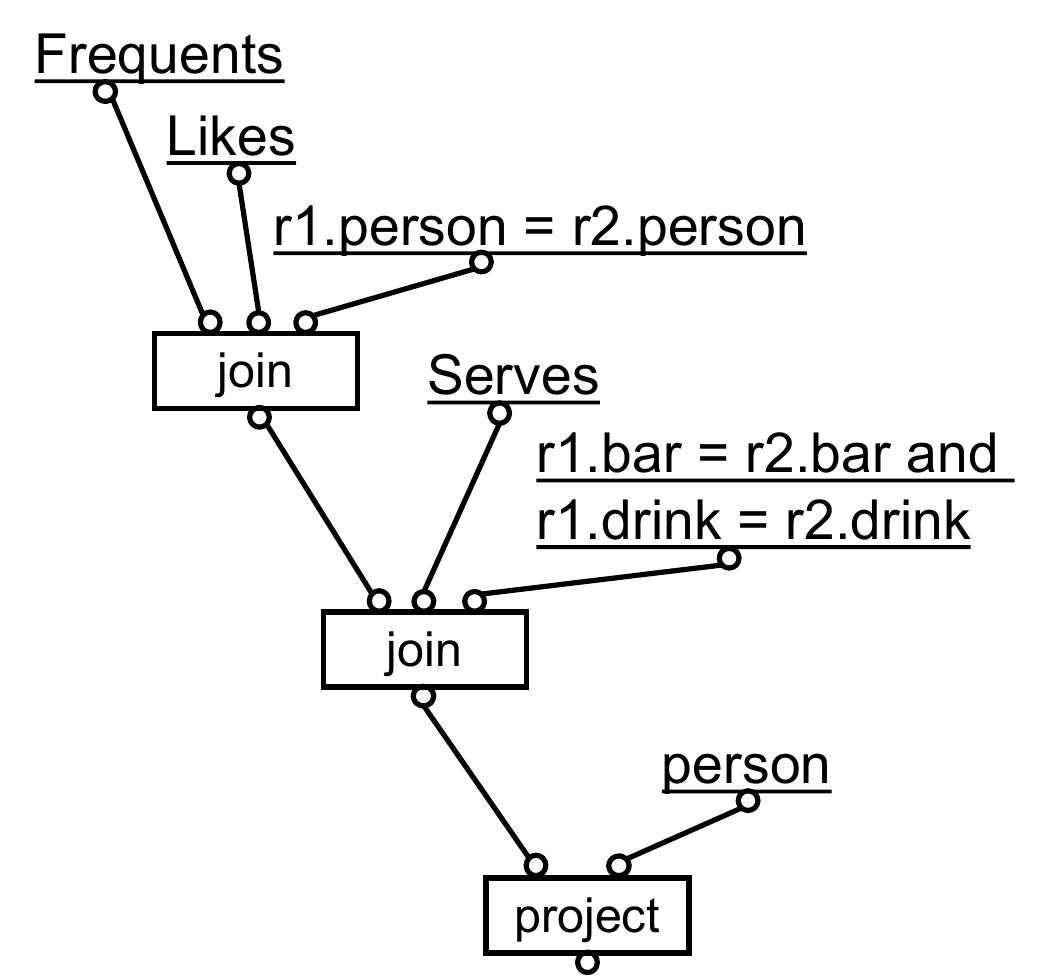}
	    \label{Fig_DFQL_Drinkers}	
	}
    \caption{
	Query visualizations are not query plans nor data flow diagrams:
	(a) Visualized query plan by Postgres' EXPLAIN command~\cite{postgres} for query $Q_{\textrm{some}}$ from \autoref{figure:sql_conjunctive}.
	(b) Same query expressed in DFQL (Dataflow Query Language) \cite{DBLP:journals/iam/ClarkW94} which is modeled after relational algebra.
	Notice that neither visualization captures the cyclic nature of the joins 
	(see \autoref{Fig_ExampleExists}).
	``\emph{Query visualization}'' to ``\emph{query plan visualization}'' 
	is the same as
	``\emph{intend of a query} (WHAT)'' to ``\emph{execution of a query} (HOW)''.
		} 
    \label{Fig_explain_PersonBarBeer}
\end{figure}

Readers may be familiar with visualizations of query plans.
\autoref{Fig_explain_PersonBarDrink} shows a query plan chosen by PostgreSQL~\cite{postgres}
to run query $Q_{\textrm{some}}$ from \autoref{figure:sql_conjunctive} 
(\emph{Find persons who frequent some bar that serves a drink they like}).
Similarly, \autoref{Fig_DFQL_Drinkers} shows the same query expressed 
in DFQL (Dataflow Query Language) \cite{DBLP:journals/iam/ClarkW94} which is modeled after relational algebra.
Notice that neither visualization captures the cyclic nature of the joins in query $Q_{\textrm{some}}$.
A query plan visualization attempts to represent HOW a query is executed.
In contrast, a query visualization attempts to represent WHAT a query does (i.e.\ its intent) and possibly the relational pattern it uses.
See the query visualization in \autoref{Fig_ExampleExists} which shows the join pattern and that this query is cyclic.
Similarly, query visualizations are also different from 
visualizing and comparing the cost or speed of execution plans~\cite{DBLP:journals/pvldb/Haritsa10}.

\subsection{Query Visualization is only partially related to Visal Query Debugging} 
\label{sec:querydebugging}

An important reason for why we want to help users understand HOW exactly a given query is executed
is for debugging a faulty query.
Visualizing software execution behavior
can be helpful for program debugging
\cite{DBLP:conf/chi/GathaniLB20,DBLP:journals/vlc/Reiss07},
but only if it helps explain WHY a query returns a particular result or WHY NOT~\cite{DBLP:conf/mud/MeliouGMS10}.
To achieve this more fine-grained understanding, state-of-the-art workflows for debugging SQL queries help users understand 
queries by somehow \emph{showing intermediate results}~\cite{
GrustKRS2011:SQLdebugging,
DBLP:journals/tods/GrustR13,
MRJ:ExplainingWrongQueries:2019,
DBLP:conf/sigmod/OlstonCS09}.
Thus it is helpful to see debugging as a spectrum of goals with different ``granularities.''
At one end of the spectrum, a query may be faulty because two incorrect tables are joined 
(a tool that answers WHAT the query actually does would help here).
At the other more-fine grained level, a query may be faulty because a DMBS implemented a particular SQL syntax for handling NULLS incorrectly\footnote{As example see \url{https://stackoverflow.com/questions/19686262/query-featuring-outer-joins-behaves-differently-in-oracle-12c}}, 
and it seems there is no way to avoid using data examples for effective debugging.

\subsection{Alternatives to Query Visualization for helping users understand existing queries} 
\label{section:alternatives}

There are three main alternative approaches for helping users understand existing queries:

\textbf{(1) Illustrating queries by examples}. 
Several papers suggest illustrating the semantics of operators in a data flow program or the semantics of queries by generating 
\emph{example input and output data}, and possibly \emph{intermediate data}.
The result is basically a list of tuples for each relational operator~\cite{DBLP:conf/uist/AbouziedHS12,
DBLP:journals/vldb/ChandraCKRS015,
GrustKRS2011:SQLdebugging,
DBLP:journals/tods/GrustR13,
DBLP:journals/pvldb/KhanX0H17,
MRJ:ExplainingWrongQueries:2019,
DBLP:conf/sigmod/OlstonCS09}.

\textbf{(2) Translating queries into Natural Language (NL)}.
Translating between SQL and NL is a heavily researched topic, and various ideas are proposed to explain queries in NL~\cite{DBLP:conf/inlg/GehrmannDER18,
DBLP:conf/nldb/Ioannidis08, 
DBLP:conf/icde/KoutrikaSI10,
DBLP:conf/cidr/SimitsisI09, 
DBLP:conf/emnlp/XuWWFS18}. 
Work in this area convincingly argues that automatically creating effective free-flowing text from queries is difficult and that the overall task is quite different from previous work on creating NL interfaces to DBMSs~\cite{Jarke:NL:1985}.
There is also recent work on translating query plans into NL~\cite{DBLP:conf/sigmod/WangBLJLC21}.

\textbf{(3) Pretty printing queries}.
Query editors for major DBMSs use 
\emph{syntax highlighting
and aligning} 
of query blocks and clauses. 
Pretty printers, such as sqlparse~\cite{sqlparse}, automatically arrange a SQL query in a supposedly easy-to-read form.
The most important dimensions are colors, capital vs.\ small letters, and indentation.

A key difference of alternatives to query visualization is 
that they are all inherently linear.
A list of tuples or a textual description do not readily reveal common logical pattern behind queries.
In particular, we are not aware of any SQL to NL tool available today that could translate our example from 
\autoref{fig:beerquery}
into an intuitive NL representation.
Patterns are naturally best shown visually,
and even the programming design patterns book~\cite{Gamma:1995ys} illustrates its patterns with intuitive diagrams.
A theory of relational query patterns, and a query-user interaction pattern inspired by ``mix-and-match'',
seems naturally supported by a visual approach.
In addition, our recent work \cite{relationalDiagrams} has shown that certain types of relational patterns 
cannot be represented in an operator-style (thus sequential) model.

\section{Principles of Query Visualization and Design trade-offs} 
\label{sec:2}

The challenge of query visualization is to find appropriate visual metaphors 
that ($i$) allow users to quickly understand a query's intent, even for complex queries,
($ii$) can be easily learned by users,
and ($iii$) can be obtained from SQL by automatic translation, including a visually-appealing automatic arrangement of nodes of the visualization.
We believe that---with the right visual alphabet---users can learn to interpret visualized queries by seeing examples without much active focus. This is similar to what is known in language learning theory as the difference between the active and the generally larger passive vocabulary: Actively reproducing newly learned content is generally more difficult than passively recognizing such content.

We next discuss the principles that led us to a particular design of a query visualization language (actually two variants, which we discuss later in more detail). We list those here to spark a healthy debate. Not all listed principles are universal, and deviations may lead to interesting alternative design decisions. These principles are also not MECE (Mutually Exclusive and Collectively Exhaustive), and some design decisions can be justified separately from other overlapping decisions.

\begin{figure}[tb]
\centering
\begin{bsubfloatrows}[\hspace{20mm}]
\bsubfloat[$Q_{\textrm{some}}$ in SQL]{%
	\textup{\textsf{
	\footnotesize
	\setlength{\tabcolsep}{1mm}
	\begin{tabular}[t]{@{} l l l l l}
			& \\
			& \textcolor{blue}{select distinct} F.person\\
			& \textcolor{blue}{from}	Frequents F, Likes L, Serves S\\
			& \textcolor{blue}{where}	F.person = L.person \\ 
			& \textcolor{blue}{and}		F.bar = S.bar \\ 					
			& \textcolor{blue}{and}		L.drink = S.drink
	\end{tabular}
	}}	
\label{figure:sql_conjunctive}	
}
\bsubfloat[$Q_{\textrm{some}}$ in \queryvis]{%
    \includegraphics[scale=0.4]{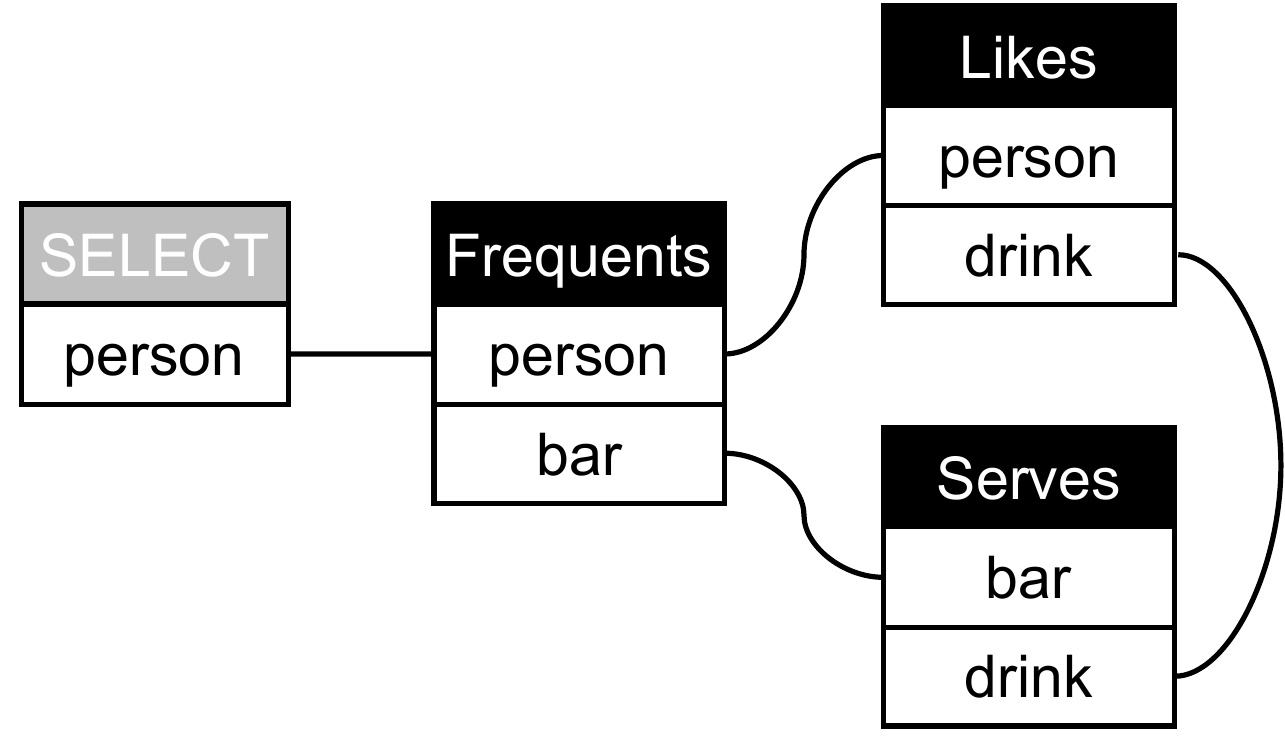}
    \label{Fig_ExampleExists}
}
\end{bsubfloatrows}
\caption{Principles 1 \& 2:
Visualizing a conjunctive query should follow a familiar UML notation:
\emph{Find persons who frequent some bar that serves some drink they like}. 
The only novelty is a dedicated output table on the left,
emphasizing the compositionality of the relational model, 
and supporting an output-oriented reading order.
}
\end{figure}

\textbf{(1) Existing metaphors as starting point}: 
Ideally, a query visualization can be learned ``on-the-fly'' by seeing visualizations of increasing complexity, starting from examples that are already familiar.
Most database users are familiar with the UML diagram notation for classes and their attributes~\cite{folwer:UMLdistilled:2003}
applied to database schemas: Table names on top of column names in rectangular bounding boxes, 
primary-foreign keys contraints represented by lines between column names.
The visualization of a conjunctive query should thus not depart too much from 
such deeply familiar visual metaphors
(e.g., see the conjunctive query $Q_{\textrm{some}}$ in \autoref{figure:sql_conjunctive}
and its visualization in \autoref{Fig_ExampleExists}).
More complicated queries then progressively extend such familiar visual metaphors.

\begin{figure}[t]
\centering
 \parbox{.34\textwidth}{	
    \subfloat[$Q_{\textrm{only}}$]{
		\textup{\textsf{
		\footnotesize
		\setlength{\tabcolsep}{1mm}
		\begin{tabular}{@{} l l l l l}
		\\[17mm]			
			& \multicolumn{4}{l}{\textcolor{blue}{select distinct} F.person}\\
			& \multicolumn{4}{l}{\textcolor{blue}{from}		Frequents F}\\
			& \multicolumn{4}{l}{\textcolor{blue}{where}	not exists} \\ 
			& \phantom{xx}	& \multicolumn{3}{l}{(\textcolor{blue}{select} 	*}\\
			& 	 			& \multicolumn{3}{l}{\textcolor{blue}{from}		Serves S}\\
			& 	 			& \multicolumn{3}{l}{\textcolor{blue}{where}	S.bar = F.bar}\\
			& 	 			& \multicolumn{3}{l}{\textcolor{blue}{and}		not exists}\\
			&				& \phantom{xx}	& (\textcolor{blue}{select} 	L.drink \\
			&				& 				& \textcolor{blue}{from}	Likes L \\
			&				& 				& \textcolor{blue}{where} 	L.person = F.person \\
			&				& 				& \textcolor{blue}{and} 	S.drink = L.drink))		
		\\[15mm]
		\end{tabular}
		}}	
		\label{figure:sql_nested}
    }
}
 \parbox{.63\textwidth}{	
    \subfloat[$Q_{\textrm{only}}$]{
        \includegraphics[scale=0.4]{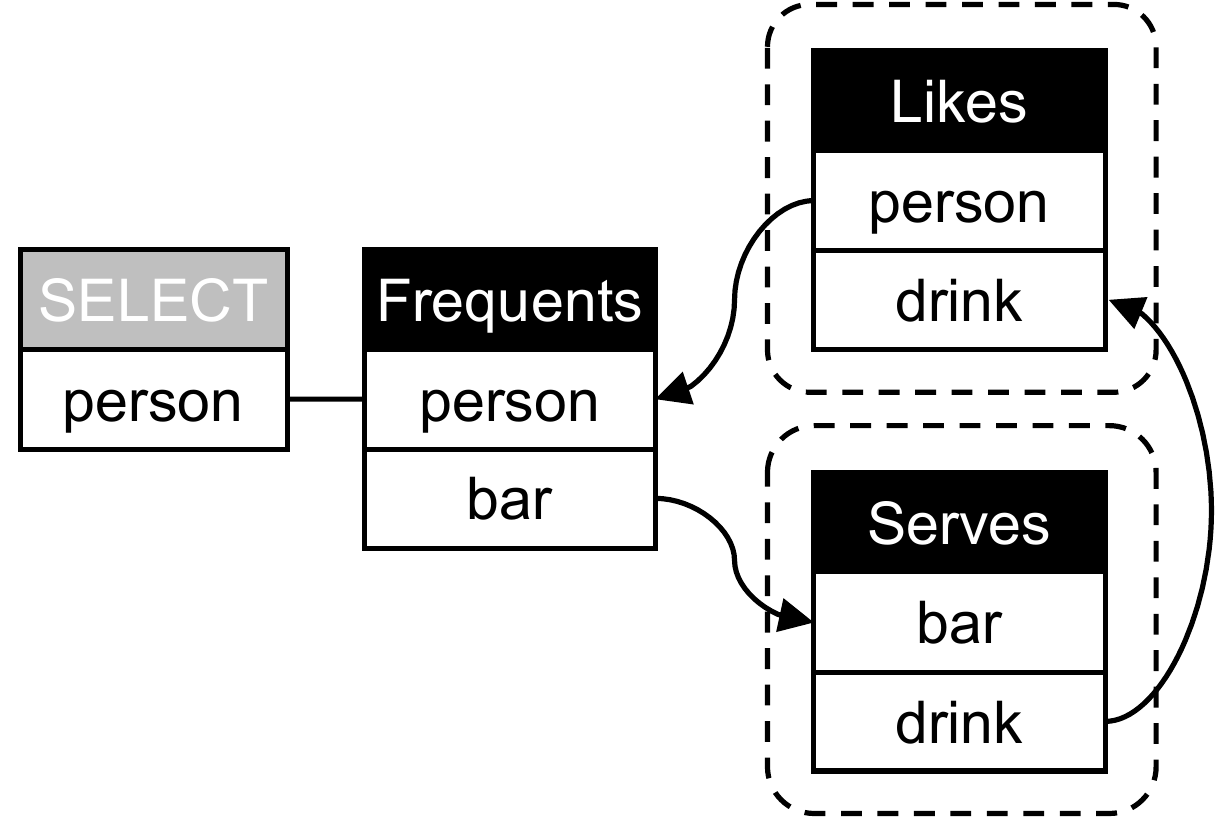}
	}
    \hspace{1mm}
    \subfloat[$Q_{\textrm{only}}$]{
        \includegraphics[scale=0.4]{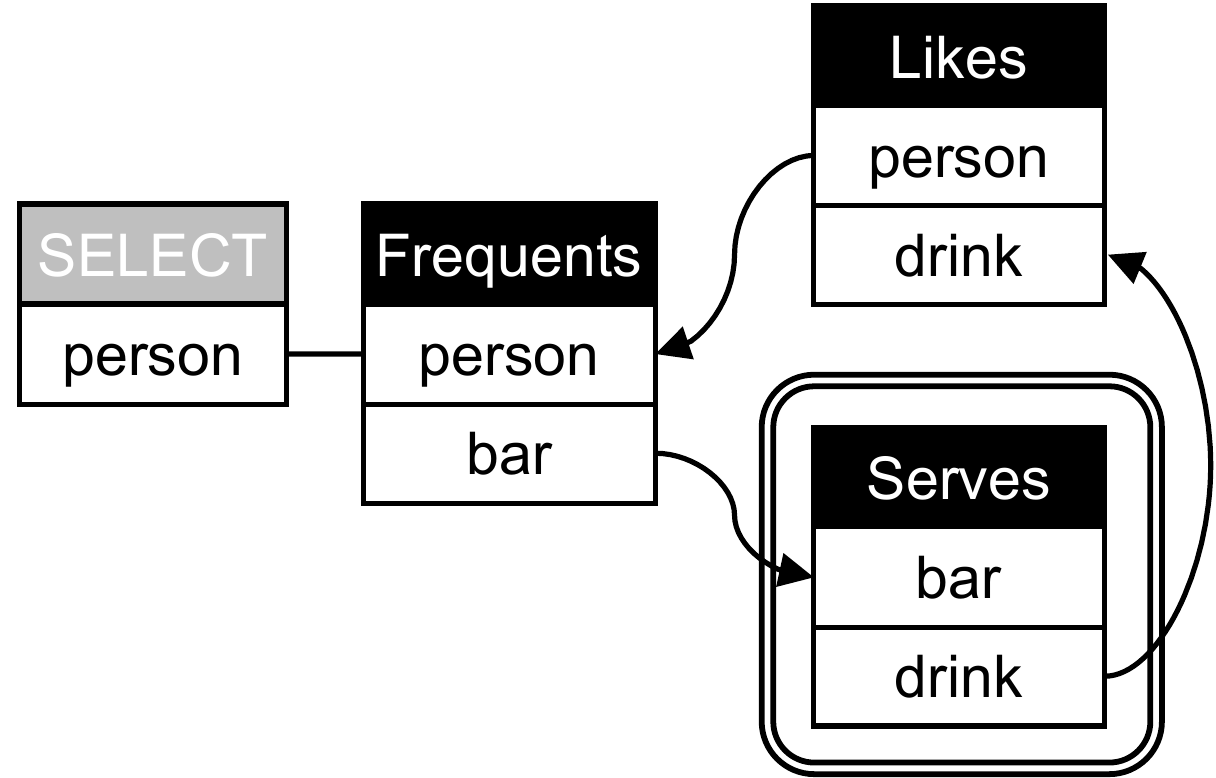}
        \label{Fig_ExampleAll}
	}
    \hspace{1mm}
    \subfloat[$Q_{\textrm{only}}$]{
        \includegraphics[scale=0.4]{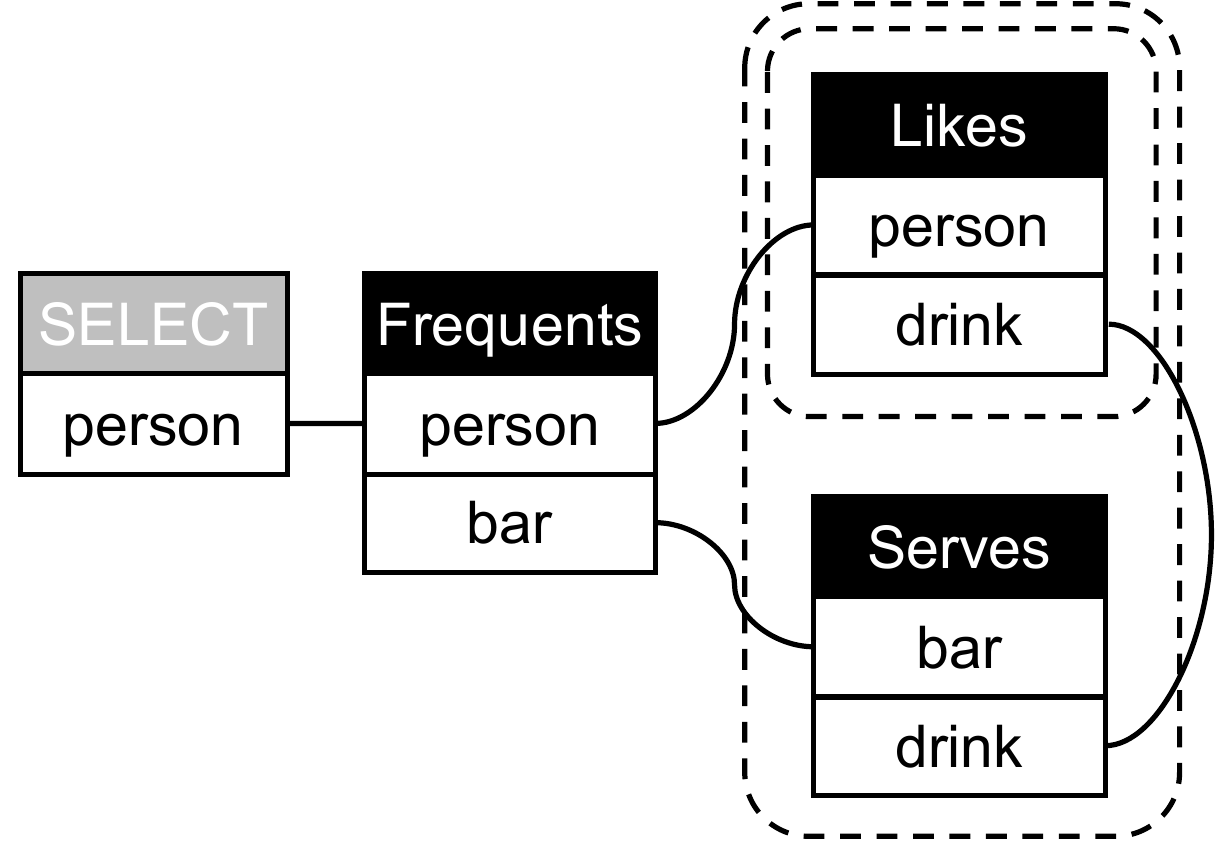}
        \vspace{1mm}
        \label{Fig_ExampleNotExistsRD}
	}
    \hspace{4mm}	
    \subfloat[$Q_{\textrm{only}}$]{
        \includegraphics[scale=0.4]{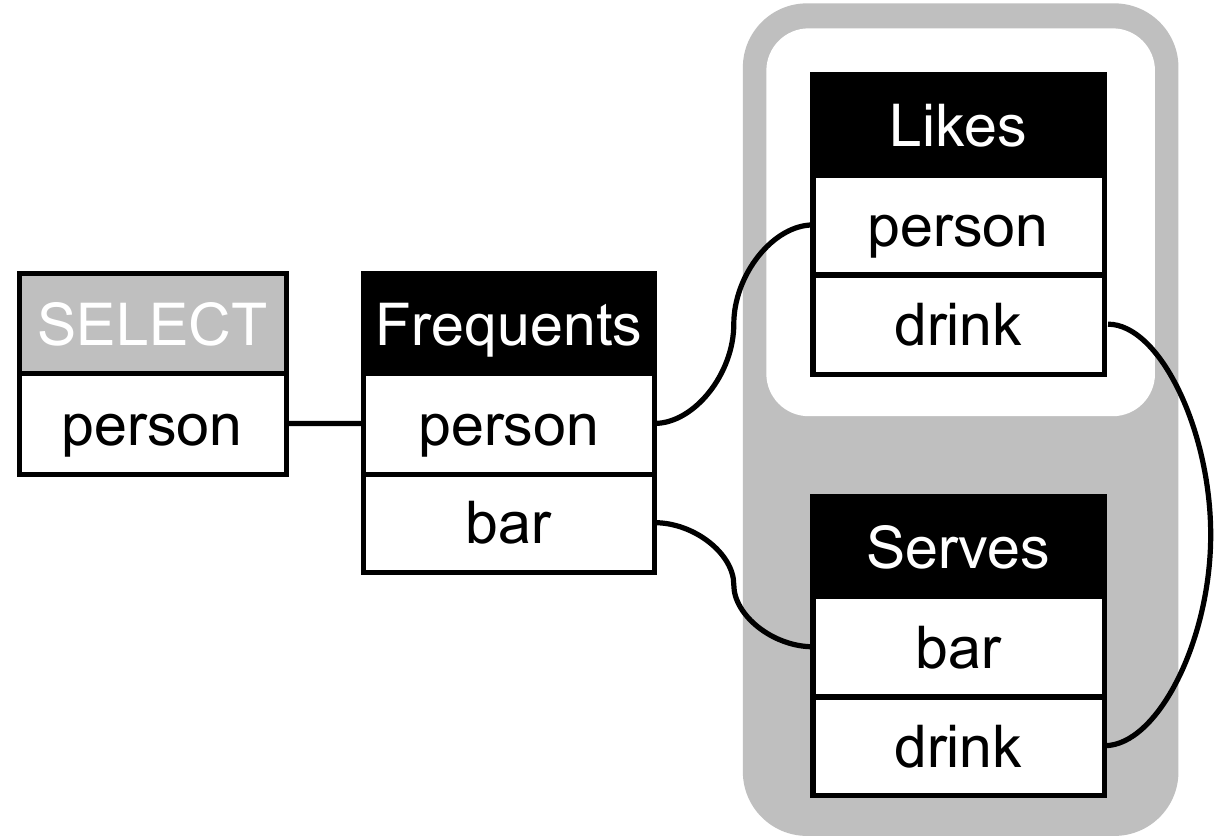}
        \vspace{1mm}
        \label{Fig_ExampleNotExistsRDShaded}
	}		
}      
    \caption{
	Principles 3 \& 8:
	(a) \emph{Find persons who frequent some bar that serves 
	ONLY drinks they like}. 
	\queryvis:
	(b) Visualizing a nested query still follows familiar UML notations, 
	but now adds visual metaphors for $\nexists$ (dashed box) and 
	the reading order can be found by following the arrows.
	(c) The reading can be further simplified by the use of the $\forall$ quantifier 
	(double-lined bounding box), a logical and intuitive operator that does not exist in SQL.
	The visualization asks for \emph{persons who frequent some bar so that 
	{ALL} drinks served are liked by them}.
	\diagrams are an alternative visualization that replaces arrows and reading orders by 
	explicit enclosure to express nesting relationships (d) and (e).
	}
    \label{Fig_ExampleVisualizations}
\end{figure}

\textbf{(2) Compositionality of the relational model}:
Inputs to queries are tables, and the output of a query is another table. 
Visualizations can (and we think should) emphasize this compositionality by explicitly showing an output table. 
This compositionality is also illustrated by the Relational Tuple Calculus expression for \autoref{figure:sql_conjunctive}:
\begin{align*}
	\{ q(\textit{person}) \mid \;
	& \exists f \in \textit{Frequents}, \exists l \in \textit{Likes}, \exists s \in \textit{Serves} 
	[
	q.\textit{person} = f.\textit{person} \wedge \\
	& 	
	f.\textit{person} = l.\textit{person} \wedge 
	l.\textit{drink} = s.\textit{drink} \wedge 	
	s.\textit{bar} = f.\textit{bar}
	] \}	
\end{align*}
The expression makes use of 4 tables: 3 input tables (\textit{Frequents}, \textit{Likes}, and \textit{Serves}), and 1 output table called~$q$
(\autoref{Fig_ExampleExists} names the output table ``SELECT'').
In contrast, all interactive query tools listed in \autoref{sec:relatedWork} use either checkmarks, stars, or colors
to highlight a subset of attributes that are returned by the query.

\textbf{(3) Progressive visual complexity}: 
Entropy codes, such as Huffman codes~\cite{cover:thomas:2006}, 
compress data by encoding symbols with an amount of bits inversely proportional to the frequency of the symbols.
In the same spirit, a visual alphabet should be adapted to an overall expected workload and visual constructs for more common logical operators should be designed with lower visual complexity than less common ones. 
Starting from UML and its familiar notations for schemas and conjunctive queries, 
we can then enhance the visual representation in a \emph{progressive way}. 
For example, almost all database queries use the logical AND in their first-order logic translation (e.g.\ joins, \sql{EXISTS}, \sql{IN}), but only few use OR (e.g.\ \sql{OR}, \sql{UNION}). 
If infrequent query constructs become increasingly complex to read, this progression does not decrease the overall usability, but rather assures that more often used constructs are simple to read, in turn.
For example, the visualization of the query from \autoref{figure:sql_nested}
is expected to be at least as ``complicated'' as the query from \autoref{figure:sql_conjunctive}.

For increasing complexity of nested queries with negation, we are inspired by a body of work on \emph{diagrammatic reasoning systems}~\cite{DBLP:conf/iccs/Howse08}. 
Diagrammatic notations are in turn inspired by the influential \emph{existential graph notation} by Charles Sanders Peirce~\cite{peirce:1933,Roberts:1992,Shin:2002}.
These graphs exploit topological properties, such as enclosure, to represent logical expressions and set-theoretic relationships
(see description in \autoref{Fig_ExampleVisualizations}).

\begin{figure}[t]
\centering
\subfloat[SQL query]{
	\includegraphics[scale=0.4]{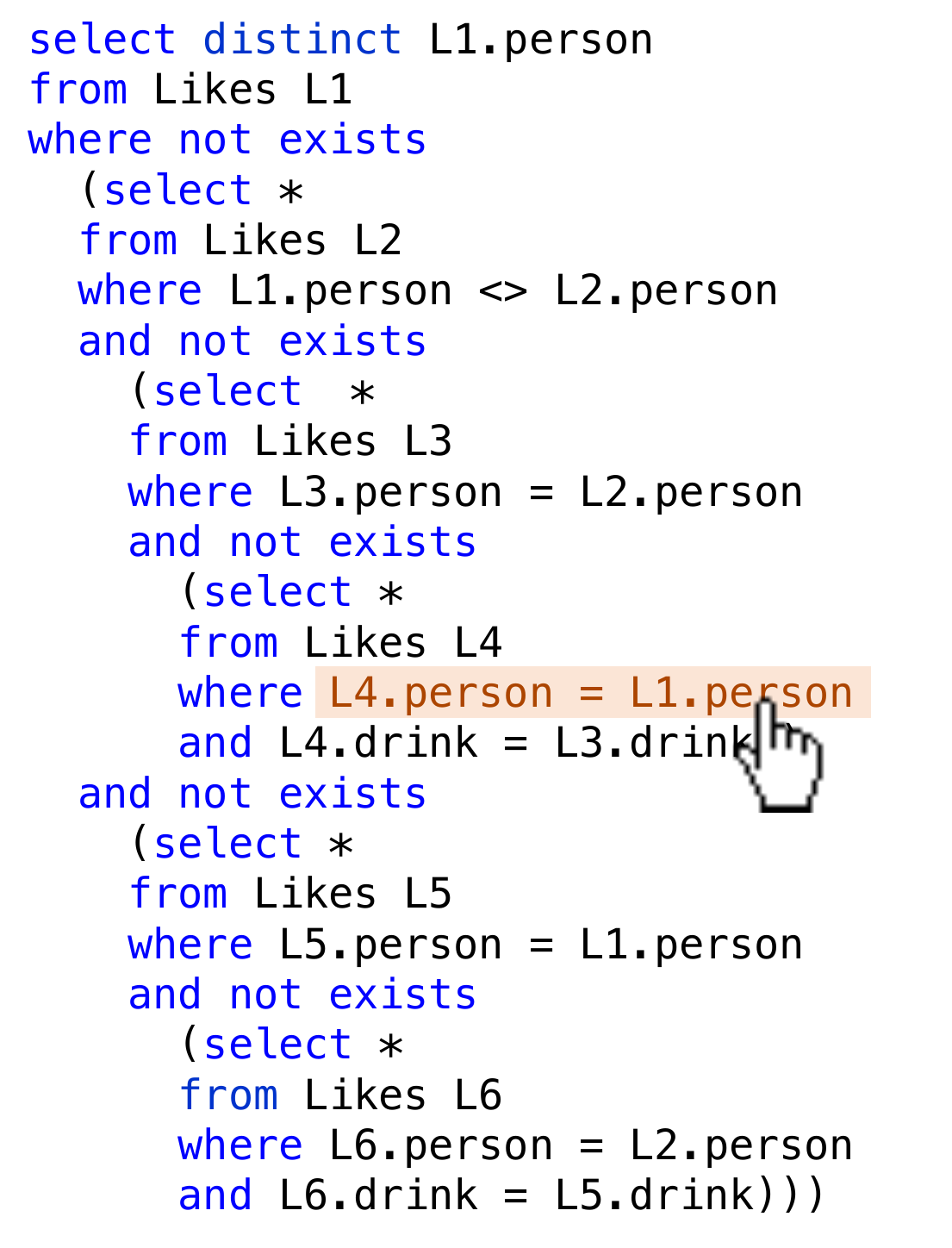}
	\label{Fig_uniqueDrinkerTastSQL}}
\hspace{10mm}
\subfloat[Query Visualization]{
	\includegraphics[scale=0.45]{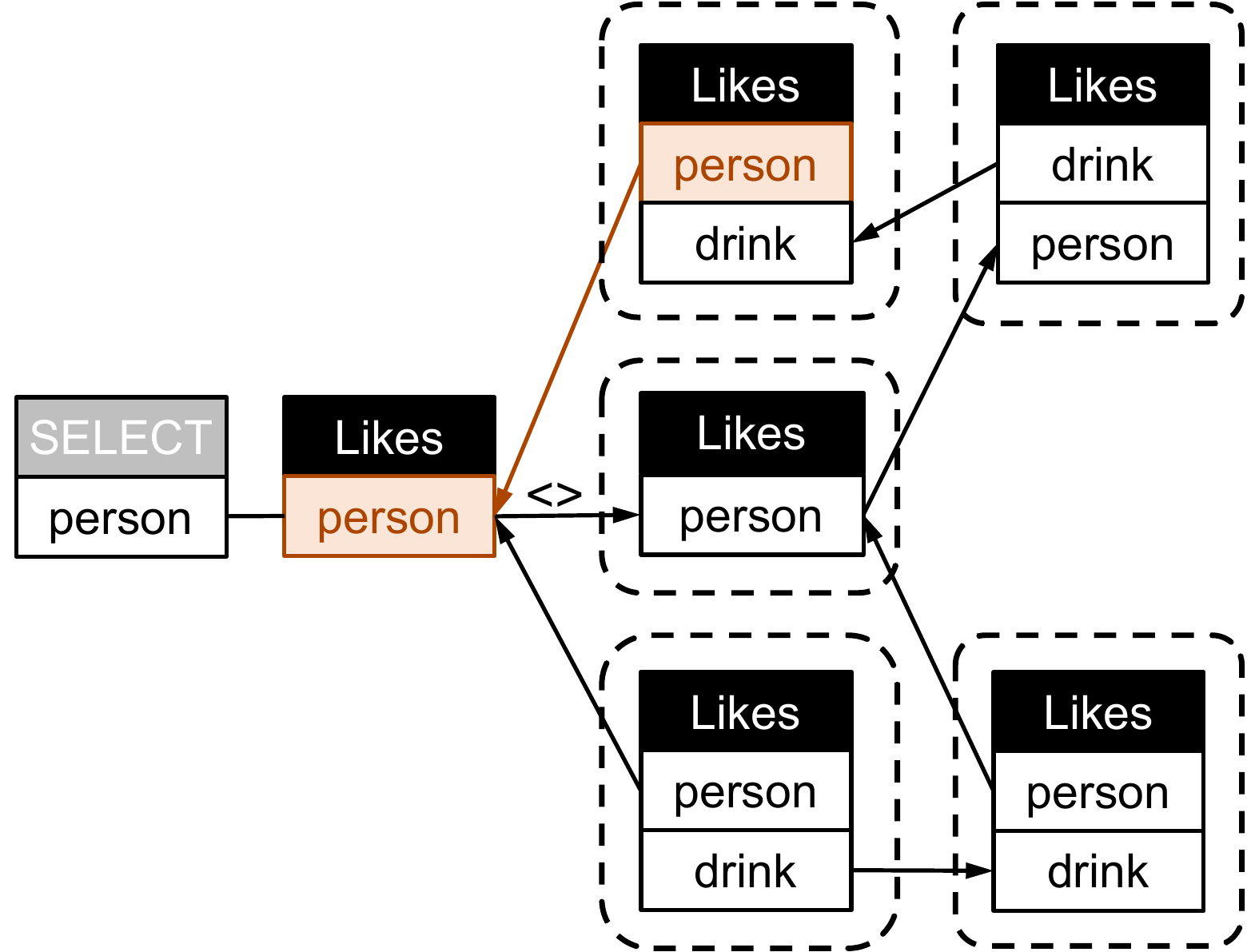}
	\label{Fig_uniqueDrinkerTastQV1}}	
\caption{Principles 4 \& 5: 
(a) Unique-set-query ``\emph{Find person with a unique drink taste}.''
(b)~\queryvis diagram with reading order encoded by arrows 
(please see \cite{DBLP:conf/sigmod/LeventidisZDGJR20} for a detailed discussion of the query and its visualization).
There is a 1-to-1 correspondence between the SQL query and its visualization. 
As the user moves the mouse over fragments of the query, the graphical representation highlights the corresponding visual elements.
}
\label{fig:beerquery}
\end{figure}

\textbf{(4) Expose (and not hide) relational patterns}:
We believe that a query visualization should expose the relational pattern used in a textual query,
instead of replacing it with an abstraction and concepts that go beyond the relational model.
This requires a visualization to use the same number of input tables of the textual query 
and to preserve a 1-to-1 mapping between them.
To illustrate, consider the SQL query in
\autoref{Fig_uniqueDrinkerTastSQL} asking for ``persons with a unique drink taste.''
The query uses 6 instances of the same table in a pattern that reads
``return any person, s.t.\ there does not exist any other person, s.t. there does not exist any drink liked by that other person 
that is not also liked by the returned person and there does not exist any drink liked by the returned person that is not also liked by the same other person.''
The visualization in \autoref{Fig_uniqueDrinkerTastQV1}
does not replace that relational pattern with another shorter construct, but rather 
makes it easier to inspect and reason about:
it complements the textual query
and preserves some traceable mapping between query and visualization. It preserves its relational pattern.

\textbf{(5) Minimal visual complexity}: 
A query visualization should fulfill some kind of minimality criteria.
Intuitively, we aim to minimize the ink-data ratio 
(thus we like to maximize its inverse: Edward Tufte's famous data-ink ratio defined as the proportion of a graphic’s “ink” devoted to the informative and thus non-redundant display of data information~\cite{tufte2001visual}).
Minimality can be interpreted in different ways:
For example, the visual alphabet could contain only a minimal set of different visual elements; removing an element would then render the visualization less expressive.
Or, for a given query, removing a particular visual element would render the query incomplete.
To achieve such minimality, one can take inspiration by comparative analysis of existing textual languages.
For example, 
($i$) Datalog does not require the use of table aliases (different occurrences of the same input relation can be distinguished by their join patterns), whereas SQL requires alias.
On the other hand, ($ii$) SQL (inspired by tuple relational calculus) does not reference any attribute that is not used by the query,
whereas Datalog uses positional information and thus requires to maintain positional information.
\autoref{Fig_uniqueDrinkerTastQV1}
shows an example \queryvis visualization that combines the best of both worlds: 
($i$) repeated relations do not require aliases, 
and ($ii$) each of those occurrences only displays attributes needed.\footnote{A visualization could still show aliases to make it easier to maintain a static correspondence between the query and it visualization. However those aliases are not needed to interpret the meaning of a visual diagram.}

\begin{figure}[tb]
\centering
\begin{bsubfloatrows}[\hspace{20mm}]
\bsubfloat[]{%
	\textup{\textsf{
	\footnotesize
	\setlength{\tabcolsep}{1mm}
	\begin{tabular}{@{} l l l l l}
		& \multicolumn{4}{l}{\textcolor{blue}{select distinct} A}\\
		& \multicolumn{4}{l}{\textcolor{blue}{from}		R}\\
		& \multicolumn{4}{l}{\textcolor{blue}{where}	B not in} \\ 
		& \phantom{xx}	& \multicolumn{3}{l}{(\textcolor{blue}{select} 	C}\\
		& 	 			& \multicolumn{3}{l}{\textcolor{blue}{from}		S)}
	\\[23mm]
	\\
	\end{tabular}
	}}	
\label{Fig_IntroductionExample_b_table}	
}
\bsubfloat[]{%
	\textup{\textsf{
	\footnotesize
	\setlength{\tabcolsep}{1mm}
	\begin{tabular}{@{} l l l l l}
		& \multicolumn{4}{l}{\textcolor{blue}{select distinct} A}\\
		& \multicolumn{4}{l}{\textcolor{blue}{from}		R}\\
		& \multicolumn{4}{l}{\textcolor{blue}{where}	not exists} \\ 
		& \phantom{xx}	& \multicolumn{3}{l}{(\textcolor{blue}{select} 	*}\\
		& 	 			& \multicolumn{3}{l}{\textcolor{blue}{from}		S}\\
		& 	 			& \multicolumn{3}{l}{\textcolor{blue}{where}	B=C)}		
	\\
	\end{tabular}
	}}	
\label{Fig_IntroductionExample_c_table}
}
\bsubfloat[]{%
	\begin{minipage}{45mm}	
	\vspace{5mm}
	\includegraphics[scale=0.4]{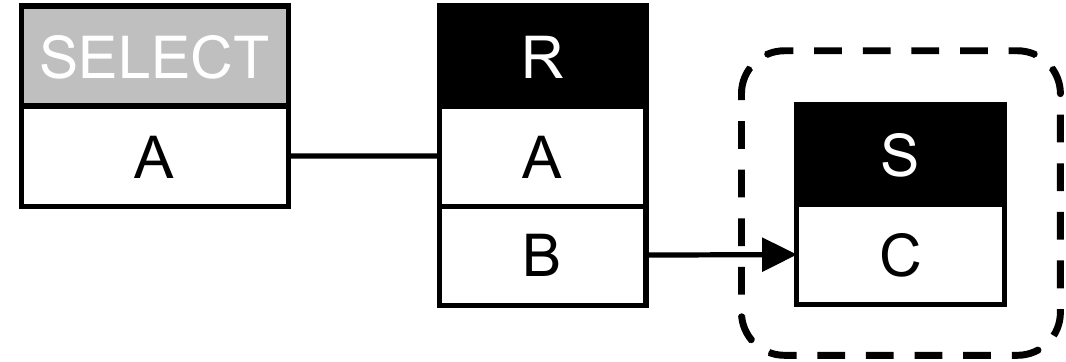}
	\end{minipage}
    \label{Fig_ExampleNotExists}
}
\end{bsubfloatrows}
\caption{Principles 6 \& 7: Queries (a) and (b) are equivalent except if column S.C contains NULL values. 
Thus ignoring NULL values in the database, they are equivalent and their \emph{query intent} can be represented by the same 
\queryvis representation shown in (c)
(example taken and slightly fixed from Fig.~4 in \cite{gatterbauer2011databases}).}
\label{Fig_CompleteExampleAmbiguous}
\end{figure}

\textbf{(6) Abstract away from syntax details}:
A query visualization abstracts away from language-specific peculiarities.
It can thus be non-injective with regard to syntactic redundancy.
A prominent example is SQL's use of NULLs. 
While there has been a lot of work on putting SQL's use of NULL values on solid foundations,
there is no universally agreed standard, and 
SQL queries evaluated on databases with NULL ``may produce answers that are just plain wrong''~\cite{DBLP:journals/sigmod/GuagliardoL17}.
The goal of query visualization can't be to provide an unambiguous interpretation of queries in the presence of NULLs, 
and thereby as a side-product also fix issues that have vexed database theoreticians over decades.
Rather, the focus of query visualization on the underlying relational patterns means that query visualizations 
need to abstract away from such oddities and not preserve them
(see \autoref{Fig_CompleteExampleAmbiguous}).
Also, a query visualization is meant to complement a textual original query.
It thus does not have to preserve all the information from the query; it can be non-injective,
thereby dividing the work: a visualization for the overall pattern, the text for the details.
This point goes back to \autoref{section:whatisQVnot} and what query visualization does not try to achieve.
The focus is on WHAT a query does, 
yet confined to the relational model and the particular underlying relational pattern
(including all the input tables used), not the syntax nor the HOW of a particular execution plan.
Tools that help users cope with the inherent syntactic difficulty of SQL fall under the category of SQL debugging (\autoref{sec:querydebugging}).
The common foundation of all relational query languages and the relational model is
First-Order Logic.
We thus believe that focusing on the \emph{logical interpretation} of queries~\cite{DBLP:journals/bsl/HalpernHIKVV01} and set semantics 
provides a solid and well-understood foundation for query visualization.
See \autoref{Fig_CompleteExampleAmbiguous} for an example.

\textbf{(7) Output-oriented reading order}:
Similar to a SQL query having an expected order of clauses (e.g.\ SELECT-FROM-WHERE),
also a query visualization benefits from having an expected arrangement and reading order.
Mirroring the design decision from SQL and calculus, we suggest a reading order left-to-right 
that starts with the result of the query (the output table) 
and then adds tables in horizontal layers in decreasing order of their relatedness to the output table (see \autoref{Fig_ExampleNotExists}). 
This suggests an arrangement where the input tables are placed at a horizontal distance from the output table that represents the shortest path join connection to the output table.
Furthermore, the arrangement of the input tables should 
be such as to simplify the reading and understanding of the query 
by following aesthetic heuristics (e.g.\ minimizing the number of line crossings).

\textbf{(8) Logic-based visual transformations}:
Nested negated quantifiers are particularly difficult to understand for users~\cite{DBLP:journals/csur/Reisner81,Reisner1975:HumanFactors}.
Simple visualizations of logical transformation can further help show the query in a more intuitive form.
Take a double negated query such as \autoref{Fig_ExampleVisualizations}:
\begin{align*}
	\{ q(\textit{person}) \mid \;
	& \exists f \in \textit{Frequents}
	[
	q.\textit{person} = f.\textit{person} \wedge
	\neg \exists s \in \textit{Serves} [s.\textit{bar} = f.\textit{bar} \wedge
	\\
	&\neg \exists l \in \textit{Likes} [
	l.\textit{drink} = s.\textit{drink} \wedge 	
	f.\textit{person} = l.\textit{person}	
	] \}	
\end{align*}
The same query can be arguably understood more easily by writing it as
\begin{align*}
	\{ q(\textit{person}) \mid \;
	& \exists f \in \textit{Frequents}
	[
	q.\textit{person} = f.\textit{person} \wedge
	\forall s \in \textit{Serves} [s.\textit{bar} = f.\textit{bar} \rightarrow
	\\
	&\exists l \in \textit{Likes} [
	l.\textit{drink} = s.\textit{drink} \wedge 	
	f.\textit{person} = l.\textit{person}	
	] \}	
\end{align*}

\section{Our Suggestions: \queryvis and \diagrams}

When following the earlier listed design principles, a family of query visualizations naturally emerges.
We discuss here two instances that differ in the way they visually encode the \emph{nesting structure between query blocks}:

(1) \queryvis~\cite{DanaparamitaG2011:QueryViz,
gatterbauer2011databases,
DBLP:conf/sigmod/LeventidisZDGJR20}:
This earlier variant from 2011
borrows the idea of a ``default reading order''
from diagrammatic reasoning systems~\cite{DBLP:conf/diagrams/FishH04} 
and uses \emph{arrows} to indicate an implicit reading order between different nesting levels.
Take as example~\autoref{Fig_ExampleNotExists} and notice how the arrows between the relations correspond to the order in which 
they appear in the natural language translation
(``Find persons who FREQUENT some bar that SERVES only drinks they LIKE''). 
Without the arrows, there would be no natural order placed on the existential quantifiers 
and the visualization would be ambiguous.
\queryvis focuses on the non-disjunctive fragment of relational calculus and is guaranteed to represent connected nested queries unambiguously up to nesting level 3.
An interactive online version is available linked from \url{https://queryvis.com}.

(2) \diagrams~\cite{relationalDiagrams}:
This more recent variant indicates the nesting structure of table variables by using \emph{nested negated bounding boxes} instead of arrows.
The nesting of negation boxes is more closely inspired by Peirce's influence beta existential graphs~\cite{peirce:1933,Roberts:1992,Shin:2002}.
Interestingly, because \diagrams are based on Tuple Relational Calculus (instead of Domain Relational Calculus which is closer to First-Order Logic)
they solve interpretation problems of beta graphs that have been the focus of intense research in the diagrammatic reasoning communities.
The big advantage of this variant is that it has a provably unambiguous interpretation for any nesting depth, 
even for queries with disconnected components,
and for both Boolean and non-Boolean queries.
Furthermore, by adding one additional visual element, \diagrams can be made relationally complete even for non-Boolean queries.
The downside is that these diagrams need more ``ink'' for simple nested queries, 
and logical transformations (design decision 8) cannot be as easily applied anymore.
An alternative to such transformations, however, is using shading for alternating nesting depths
(see e.g.\ \autoref{Fig_ExampleNotExistsRDShaded}).

An interactive online version of \queryvis has been online at \url{https://queryVis.com} since 2011~\cite{DanaparamitaG2011:QueryViz}.
We encourage the reader to try it. 
It currently supports only a limited SQL grammar (see the web page for details). 
Still, this online demo shows that query visualization can have a very lightweight interaction. 
The user does not have to specify anything upfront and can just copy the SQL query and the schema into the two available forms
(notice that the relevant part of the schema could often be inferred from the query). 

\subsection{User study showing users can interpreting queries faster with $\queryvis$}
\label{sec:userstudy}

We designed a user study to test whether our diagrams help users understand SQL queries \emph{in less time} and \emph{with fewer errors}, on average.
The study design and analysis plan was preregistered before we started the experiment and gathered data.  
Details on the study are available 
in \cite{DBLP:conf/sigmod/LeventidisZDGJR20} 
and on OSF at \osfprereg.

\begin{figure}[t]
\centering
\includegraphics[scale=0.6]{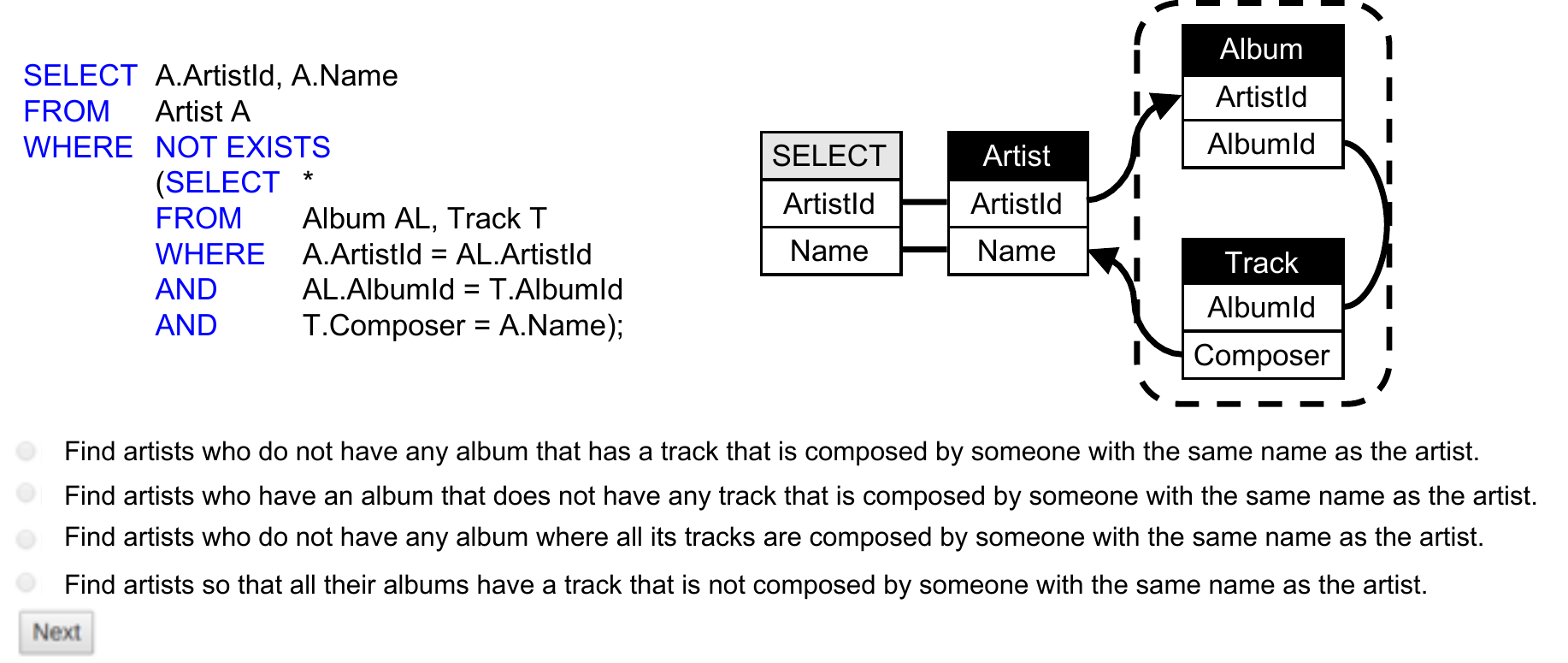} 
\caption{
	\autoref{sec:userstudy}: 
	Example query from our user study.
    The query is shown in the \emph{Both} condition, in which a participant sees the query in both SQL (left) and 
	our \queryvis diagram (right).
	}
\label{figure:study_screenshot}
\end{figure}

The study is an easily-scalable \emph{within-subjects study}~\cite{seltman2012experimental} 
(i.e., all study participants were exposed to all query interfaces).
The study consisted of 9 multiple-choice questions (MCQs).
Each MCQ asked the participant to choose the best interpretation for a presented query from 4 choices.
Following best practices in MCQ creation~\cite{Zimmaro2010},
we created all 4 choices to read very similar to each other so that a participant with little knowledge of SQL would be incapable of eliminating any of the 4 choices.
Each query was presented to participants in one of 3 conditions: 
(1) seeing a query as SQL alone (``SQL''), 
(2) seeing a query as a logical diagram that was generated from SQL (``QV''), or 
(3) seeing both SQL and \queryvis at the same time (``Both'').
\autoref{figure:study_screenshot} shows the interface for the condition ``Both'' for one of the 9 questions.
Each participant answered \emph{all 9 questions in the same order} but
the condition for each question was 
randomized in a particular way that reduces potential biases in our analysis due to condition ordering effects
following a \emph{Latin square design}~\cite{Ledolter:2007fu,Montgomery:2013wq}.
We then tracked the time needed and errors made by each participant while trying to find the correct interpretation for each query.
Participants had to pass an SQL qualification exam to ensure that they had at least a basic proficiency with SQL.
We made the study available for 3 weeks from Jan 24, 2020--Feb 13, 2020 on Amazon Mechanical Turk (AMT),
during  which we recruited $n = 42$ legitimate participants.

\begin{figure}[t]
\centering
\subfloat[]{
    \includegraphics[scale=0.44]{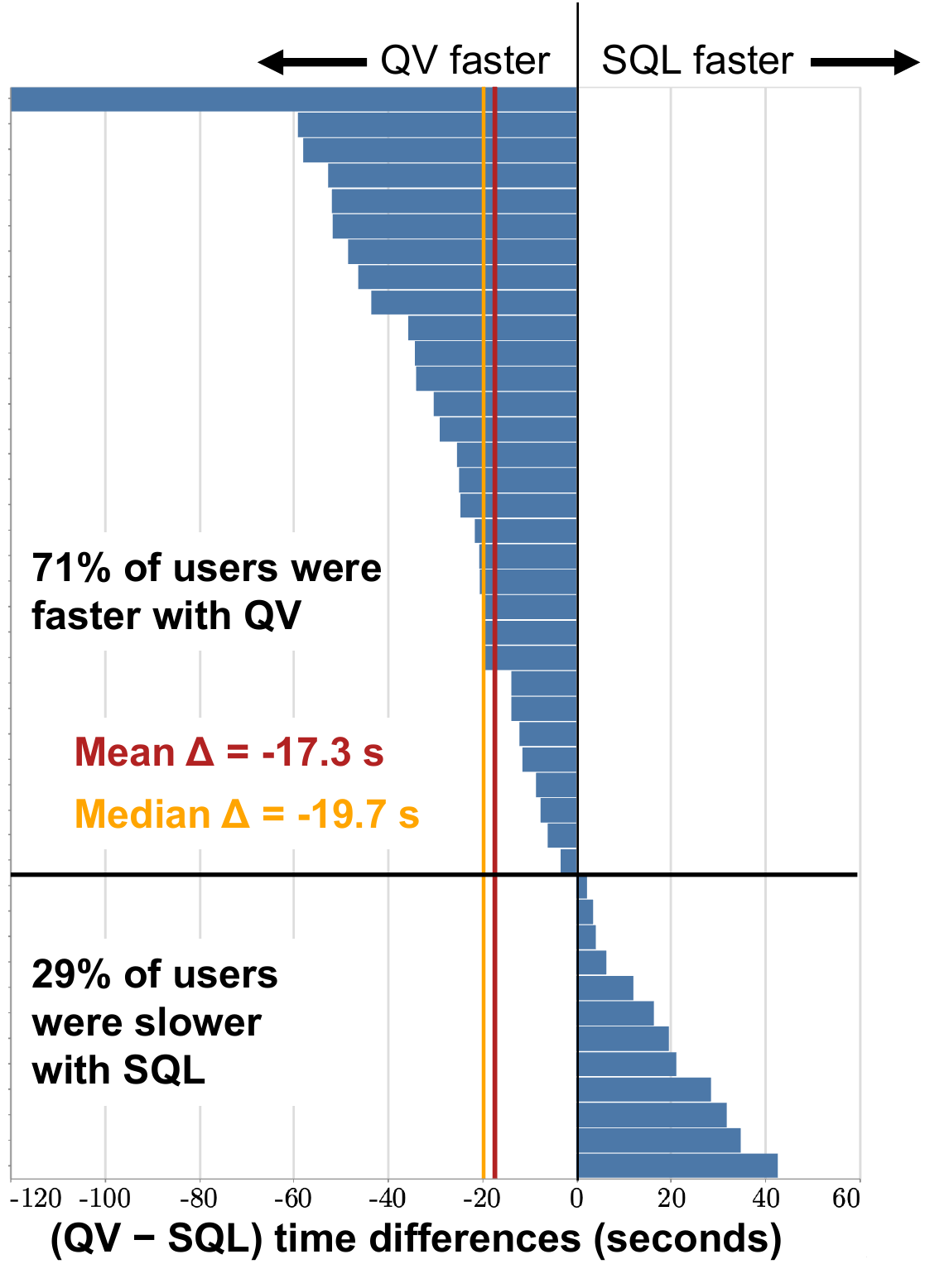}
    \label{figure:time_differences_no_grouping}
}
\hspace{25mm}
\subfloat[]{
    \includegraphics[scale=0.44]{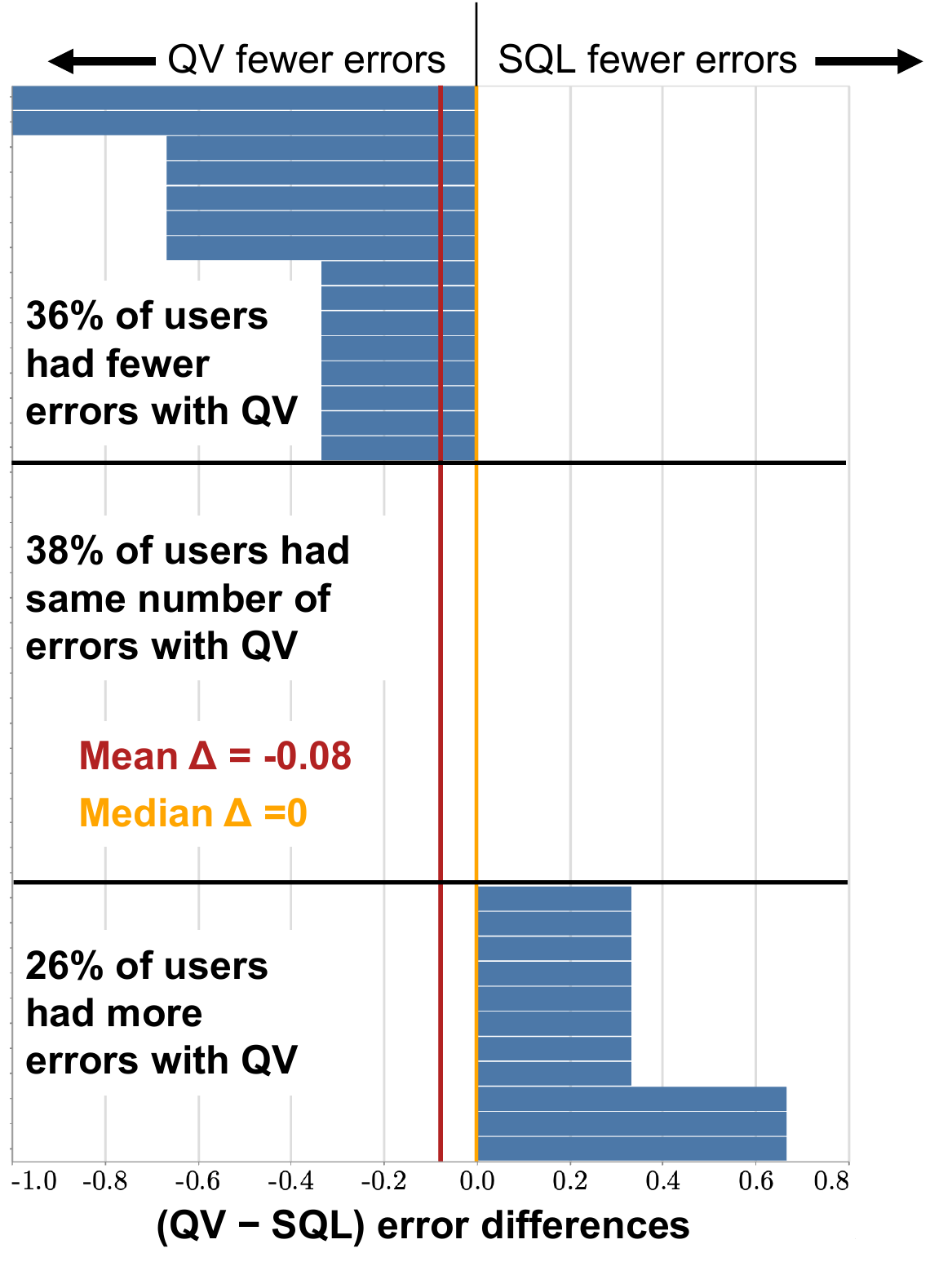}
	\label{figure:error_differences_no_grouping}
}
\caption{\autoref{sec:userstudy}:
Distribution of (QV$-$SQL) time and error differences for each participant on 9 MCQs.
}
\label{figure:differences_no_grouping}
\end{figure}

\introparagraph{Results}
There is strong evidence that participants are meaningfully faster (-20\%) using \queryvis than {SQL} 
($p<0.001$). 
There is weak evidence that participants make meaningfully fewer errors (-21\%) using \queryvis than {SQL}
($p=0.15$).
\autoref{figure:differences_no_grouping}
shows the full time and error difference distribution across the 42 participants.

\section{Related work}
\label{sec:relatedWork}

For decades, SQL has been the main standard for issuing queries over relational databases.
There is a reasonable chance that this widely implemented interface won't be replaced anytime soon. 
Thus we do not propose new ways to write queries, 
but instead explore how to help users \emph{understand existing SQL queries}.

\introparagraph{Visual Query Languages (VQL)}
Visual methods for expressing queries have been studied
extensively in the database literature~\cite{DBLP:journals/vlc/CatarciCLB97}, 
and
many commercial database products offer some visual interface for users to write simple SQL queries.
Query Visualization (QV) focuses on the problem of describing and \emph{interpreting a query that has already been written}, which is 
different from the problem of composing a new query (\autoref{QVisnotVQL}).

\textbf{Interactive query builders} employ 
visual diagrams that 
users can manipulate (most often in order to select tables and attributes)
while using \emph{a separate query configurator}
(similar to QBE's condition boxes~\cite{DBLP:journals/ibmsj/Zloof77}) 
to specify selection predicates, attributes, and sometimes nesting between queries.
dbForge~\cite{dbforge} is the most advanced and commercially supported tool we found for interactive query building.
Yet it does not show any visual indication for non-equi joins between tables 
and the actual filtering values and aggregation functions can only be added in a separate query configurator.
Moreover, it has limited support for nested queries: 
the inner and outer queries are built separately,
and the diagram for the inner query is \emph{presented separately and disjointly} 
from the diagram for the outer query.
Thus  \emph{no visual depiction of correlated subqueries is possible}.
Other graphical SQL editors like SQL Server Management Studio (SSMS)~\cite{ssms}, Active Query Builder~\cite{activequerybuilder}, QueryScope from SQLdep~\cite{queryscope}, MS Access \cite{msAccess}, and
PostgreSQL's pgAdmin3~\cite{pgadmin} lack in even more aspects of visual query representations: 
most do not allow nested queries, 
none has a single visual element for the logical quantifiers 
\texttt{NOT} \texttt{EXISTS} or \texttt{FOR} \texttt{ALL},
and all require specifying details of the query in SQL or across several tabbed views 
\emph{separate from a visual diagram}.
DataPlay~\cite{DBLP:conf/uist/AbouziedHS12} 
allows a user to specify their query by interactively modifying a \emph{query tree with quantifiers}
and observing changes in the matching/non-matching data.
Gestural query specification~\cite{10.14778/2732240.2732247}
allows a user to query databases using a series of gestures on a touchscreen.
In short, current graphical SQL editors \emph{do not provide a single encompassing visualization of a query}.
Thus they could not (even in theory) transform a complicated SQL query
into a single visual representation, which is the focus of query visualization.

\textbf{Query visualizations}
attempt to create succinct visual representations of existing queries.
This explicit reverse functionality for SQL has not drawn as much attention as visual query builders,
and there are only a handful of other systems we are aware of~\cite{gatterbauer:diagrams:tutorial:2022}.
Visual SQL~\cite{DBLP:conf/er/JaakkolaT03} is 
a visual query language that also support query visualization. 
With its focus on query specification, it maintains the one-to-one correspondence to SQL,
and syntactic variants of the same query lead to different representations (\autoref{Fig_CompleteExampleAmbiguous}).
SQLVis~\cite{DBLP:conf/vl/MiedemaF21} shares motivation with \queryvis. 
Similar to Visual SQL, it places a stronger focus on the actual syntax of a SQL query 
and syntactic variants like nested EXISTS queries change the visualization.
Snowflake join~\cite{snowflakejoin} is an open source project that visualizes join queries with optional grouping.
It does not have any consistent and unambiguous notation for nested queries.
GraphSQL~\cite{DBLP:conf/dexaw/CerulloP07} uses visual metaphors that are different from typical relational schema notations 
and visualizations, even simple conjunctive queries can look unfamiliar.
The Query Graph Model (QGM) developed for Starburst~\cite{DBLP:conf/sigmod/HaasFLP89}
helps users understand query plans, not query intent (\autoref{sec:queryplans}).
StreamTrace \cite{DBLP:conf/chi/BattleFDBCG16}
focuses on visualizing temporal queries
with
workflow diagrams and a timeline.
It is an example of visualizations for spatio-temporal domains and not the logic behind general relational queries.

\section{Various Challenges}

A vision paper that some of us wrote in 2011 declared that ``Databases will visualize queries too''~\cite{gatterbauer2011databases}.
This has not yet widely happened.
Why so? Is it just a matter of time?  Or is it that we are missing something profound and Queries Visualizations (QV) will go the same route as Visual Query Languages (VQL):
intuitively attractive, but practically not as useful?
Instead, perhaps the foundations have been laid, yet there are still problems to be solved
to make that vision practical.
Here is a partial list of several such challenges:

\textbf{(1) Extensions}:
Expressing logical disjunction in diagrams is inherently more complicated than conjunction~\cite{Shin:2002}.
And while relational algebra, relational calculus, and Datalog all use set semantics, SQL uses bag semantics.
What are \emph{appropriate visual metaphors} for 
general disjunctions 
and non-logical constructs, such as 
groupings, aggregates, arithmetic predicates, bag semantics, outer joins, null values, and recursion?\footnote{The online \queryvis interface at \url{https://queryVis.com} has been quietly supporting limited forms of aggregates already since 2016.}

\textbf{(2) Relational patterns}:
Something not well understood or even formalized today is the vague concept of ``relational query patterns.''
What are query patterns?
We posit that identifying patterns in queries may have several advantages, 
akin to how formalizing best practices in software design patterns has aided software engineers \cite{Gamma:1995ys}.
General and reusable query patterns could assist in teaching students how to write complicated queries.
Queries written using common patterns could then potentially be easier to interpret quickly.
What is \emph{a rigorous semantic definition of relational query patterns}?
See \cite{relationalDiagrams} for first steps in that direction.

\textbf{(3) Measures of visual conciseness}: 
We listed minimal visual complexity as guiding principle. 
When comparing two languages one could likely develop metrics such as amount of ink used.
However, what is the ultimately right measure for quantifying visual complexity for a human user?
Visual complexity also needs to take into account prior familiar notions (like UML) to a target audience.

\textbf{(4) Automatic layout algorithms}:
The online \queryvis demo uses a layered arrangement of tables that guarantees that any join conditions
are either between two adjacent layers or within the same layer.
It uses the standard Graphviz library for arranging the tables and their attributes~\cite{Ellson03graphvizand}.
However, existing graph layout algorithms are not suited for complicated layered graphs with nested hierarchies.
What are new outline algorithms that can optimize for existing visual metrics (such as minimum line overlap) 
and define novel metrics that capture visual homogeneity?
A possible route is 
encoding existing aesthetic heuristics (such as those in \cite[Table 1]{DBLP:conf/cae/BennettRSG07})
or novel ones
in quantitative metrics, and then
defining the layout problem as Integer Linear Program (ILP). 
A recent proposal that uses this route is STRATISFIMAL LAYOUT~\cite{DBLP:journals/tvcg/BartolomeoRGD22}.

\textbf{(5) Interactive diagrams}:
As already implied by \autoref{Fig_KevinBacon},
\autoref{Fig_TheVision}, 
\autoref{Fig_SpeechAssistant},and
\autoref{fig:beerquery},
the interaction between textual query and query visualization could be more involved beyond a simple one-way translation.
Interactive mouse-over can show correspondences. 
An interactive auto-complete feature could suggest possible query templates. 
And a user could be allowed to manipulate an existing template of a query, which then gets reflected in the text
(but notice that this last point would defeat the original idea to keep the visualization lightweight and as easy add-on to an existing query composition workflow).
What is the optimal end-to-end integration of visualization and text or alternative input forms
(recall \autoref{Fig_SpeechAssistant} and \autoref{Fig_TheVision})?

\textbf{(6) Combinations with other modalities}:
We mentioned in \autoref{section:alternatives} alternative ways to help users understand their queries.
Such alternative modalities could possibly be combined with visualizations.
For example, Natural Language translations could possibly benefit from a graphical representation of a query.
Parts of a query could be replaced with an automatically created text and expanded with a click to the full pattern.
Query visualizations could be enhanced with example database instances, or operator-by-operator translations.
The query visualizations could be modified to display how individual records fit or don’t fit the query. 
This could again be done with interactive mouseover or choice from a menu.

\textbf{(7) User studies}: 
We believe that preregistered, within-subjects studies with multiple-choice questions in Latin square design studies, 
where the correctness of a user's answer can be determined automatically,
are the way to go to for easily scalable quantitative comparison between different interfaces. 
In our user study, users needed to pass a SQL qualifying exam and then started the test only after minimum exposure to \queryvis. 
One could only imagine the improvement after the users had chance to become more familiar with the visual language over an extended period of time.
For such a longitudinal study that possibly instruments across control groups there is one big open challenge:
How to parameterize SQL queries and questions in the spirit of Gradiance
\cite{gradiance}
such that the same subjects can take the same test repeatedly?
Such new user study paradigms would allow us 
to observe the improvement in speed and accuracy over an extended period of time.

\textbf{(8) Declarative programming}:
Logical query interfaces have shown success also beyond relational data.
Examples include Datalog for networks
\cite{DBLP:conf/sigmod/HuangGL11,DBLP:journals/cacm/LooCGGHMRRS09,DBLP:conf/sosp/LooCHMRS05}
and Inductive Logic Programming~\cite{SchmidMuggleton:2017}. 
Such programs represent an explicit symbolic structure that can be inspected and understood, and the implied logic visualized.

\section{Conclusions and Future Work} 
\label{sec:conclusions_and_future_work}

We discussed the potential of query visualizations for a future, advanced user-query interaction
that visualizes relational patterns of a query with diagrams.
We delineated query visualization from visual query languages,
discussed a few principles for designing intuitive visual diagrams,
and gave two variants of a family of query visualizations.
Future work needs to  extend visual formalisms for the full relational model and beyond,
find algorithms for automatic arrangement of diagrams,
study novel and more interactive user interfaces 
with query visualizations being just one component, 
create more easily verifiable, large-scale user studies,
and find ways to apply logical representations to other logic-based languages such as Inductive Logic Programming.

\subsection*{Acknowledgements} 

This research is supported in part by NSF awards IIS-1762268, IIS-1956096, and IIS-2145382.
WG would also like to thank \href{https://www.linkedin.com/in/danaparamita/}{Jonathan Danaparamita} 
who created the original interactive QueryVis demo that has been online at 
\url{https://queryvis.com/} since 2011 and
without whom there would have been no \queryvis.

\bibliographystyle{abbrv}

\end{document}